\documentclass[a4paper,11pt]{article}
\pdfoutput=1 

\usepackage{jheppub} 

\usepackage[T1]{fontenc} 

\title{\boldmath D-brane Potentials in the Warped Resolved Conifold and Natural Inflation}

\author{Zachary Kenton}
\author{and Steven Thomas}

\affiliation{Centre for Research in String Theory\\School of Physics and Astronomy \\Queen Mary University of London\\Mile End Road, London E1 4NS, UK}

\emailAdd{z.a.kenton@qmul.ac.uk}
\emailAdd{s.thomas@qmul.ac.uk}

\abstract{In this paper we obtain a model of Natural Inflation from string theory with a Planckian decay constant. We investigate D-brane dynamics in the background of the warped resolved conifold (WRC) throat approximation of Type IIB string compactifications on Calabi-Yau manifolds. 
When we glue the throat to a compact bulk Calabi-Yau, we generate a D-brane potential which is a solution to the Laplace equation on the resolved conifold. We can exactly solve this equation, including dependence on the angular coordinates.
The solutions are valid down to the tip of the resolved conifold, which is not the case for the more commonly used deformed conifold. This allows us to exploit the effect of the warping, which is strongest at the tip. 
We inflate near the tip using an angular coordinate of a D5-brane in the WRC which has a discrete shift symmetry, and feels a cosine potential, giving us a model of Natural Inflation, from which it is possible to get a Planckian decay constant whilst maintaining control over the backreaction. This is because the decay constant for a wrapped brane contains powers of the warp factor, and so can be made large, while the wrapping parameter can be kept small enough so that backreaction is under control. }

\usepackage{braket}

\numberwithin{equation}{subsection}	

\newcommand\be{\begin{equation*}}
\newcommand\ee{\end{equation*}}
\newcommand\ben{\begin{enumerate}}
\newcommand\een{\end{enumerate}}
\newcommand\bal{\begin{align*}}
\newcommand\eal{\end{align*}}
\newcommand\bi{\begin{itemize}}
\newcommand\ei{\end{itemize}}

\def\I{\item}

\newcommand{\half}{\frac{1}{2}}

\def\id{\protect{{1 \kern-.28em {\rm l}}}}

\newcommand{\p}{\partial}

\newcommand{\gae}{\lower 2pt \hbox{$\, \buildrel {\scriptstyle >}\over {\scriptstyle
\sim}\,$}}
\newcommand{\lae}{\lower 2pt \hbox{$\, \buildrel {\scriptstyle <}\over {\scriptstyle
\sim}\,$}}

\begin{document} 
\maketitle
\flushbottom

\section{Introduction} \label{sec:Intro}

Inflation is a compelling solution to the horizon problem of the standard big bang model. A simple model involves a scalar field coupled to gravity together with a suitably flat potential, which drives the accelerated expansion of the early universe. Quantizing the scalar field driving inflation, we can generate quantum fluctuations which source the temperature anisotropies seen in the CMB. Quantizing the gravitational field (semiclassically) leads to gravitational waves which can source quadrupolar temperature anisotropies in the CMB, producing  polarization of the photons \cite{Polnarev1985},  \cite{Ade:2014gua}. 

String theory provides a consistent framework in which to investigate quantum fields coupled to gravity, and its UV completion, which has important effects on models of inflation.
There have been many reviews on inflation in string theory, for example \cite{String_Cosmo}, \cite{Baumann:2014nda}. 
Critical superstring theory has geometric solutions with a 10D spacetime, while in cosmology we are interested in a 4D spacetime. In Section~\ref{sec:sugra} we consider a compactification of the six dimensions on a three complex-dimensional Calabi-Yau manifold, which preserves one quarter of the original supersymmetries. 

In Section~\ref{sec:NI} we apply our string theory compactification to a model of Natural Inflation \cite{Freese1990,PhysRevD.47.426,Freese:2004un,Freese2014}. In contrast to many previous studies of Natural Inflation from a controlled string theory, we are able to achieve a Planckian decay constant. This is because our axion is an open string modulus, rather than a closed string axion formed from integrating a $p$-form over a $p$-cycle.

Calabi-Yau compactifications involve many closed string moduli, including complex structure moduli, K\"{a}hler moduli and the axiodilaton. It was originally hoped that one of these closed string moduli may provide a candidate scalar field for inflation. However, because there are so many of them, most will require stabilization, as multifield inflation is constrained by current observations \cite{Ade:2013uln}. 
In type IIB supergravity, the complex structure moduli and the axiodilaton can be stabilized classically via flux compactifications \cite{Giddings2002}, which involve a warped spacetime. The K\"{a}hler moduli are not stabilized classically and are instead fixed by quantum perturbative and non-perturbative effects  \cite{Kachru:2003sx}, \cite{Kachru:2003aw}.

In general it's difficult to stabilize all the closed string moduli, while maintaining a flat potential for just one or two of them. A more promising approach is to inflate using open string brane moduli, arising from spacetime-filling branes whose coordinates within the internal 6D space are moduli for the 4D effective theory \cite{1999PhLB..450...72D}. The branes feel a potential through interactions with other sources in the internal space. The warping helps to keep the brane potential flat, which is desired for inflation.

In order to study the dynamics of branes in warped spacetimes, we require the metric on the internal unwarped 6D Calabi-Yau space. However, no explicit metric is known on any global compact Calabi-Yau space. The best we can do is to approximate the Calabi-Yau by a noncompact throat region, which is Ricci flat and K\"{a}hler, and on which we know the metric. We can then cut off the throat at a finite length and glue it on to a compact bulk Calabi-Yau, on which the metric is unknown. Research is constricted to obtaining inflation from D-branes confined to the warped throat region, where the metric is known. 

The singular conifold (SC) is one example of a Ricci flat, K\"{a}hler throat on which the metric is known \cite{Candelas1990}. The SC lives at a singular point within the moduli space of Calabi-Yau manifolds. It is a cone over a $T^{1,1} \equiv [SU(2) \times SU(2)]/{U(1)}$ base, and so it contains a conical singularity at its tip (not to be confused with the singular point within the moduli space). We can smooth out this conical singularity in two topologically distinct ways, whilst preserving the Ricci flat and K\"{a}hler conditions, leading to the deformed conifold (DC) and the resolved conifold (RC). The DC and RC are both noncompact, and the metric is known explicitly in each case  \cite{Candelas1990,Minasian:1999tt,Ohta:1999we,  Klebanov:2000nc, 2000JHEP...08..052K} and  \cite{Zayas2000}. 

The DC is the usual choice when one is interested in stabilizing the closed string moduli and inflating within the \textit{same} throat, because the DC can support a non-trivial (2,1)-form flux which can stabilize the complex structure moduli and the axiodilaton while preserving supersymmetry \cite{Kachru:2003sx,Kachru:2003aw,2000NuPhB.584...69G}. 
This flux is also responsible for the warping of the full 10D solution,  known as the warped deformed conifold (WDC) \cite{2000JHEP...08..052K}. 

The RC, on the other hand can't support topologically non-trivial (2,1)-form flux. This means the complex structure moduli and axiodilaton can't be stabilized in the same manner as the DC.
Indeed, the RC on its own has no complex structure moduli to stabilize, and so it is no surprise that the flux mechanism used to stabilize the complex structure modulus of the DC is not suitable here\footnote{  The authors of \cite{ Dasgupta:2008hw} have analysed susy breaking ISD (1,2) fluxes on the RC, which are allowed if Poincare duality is
broken either through non-compactness or through having a compact but non-CY manifold. It would be interesting to see if this work can be extended to the case of the WRC which we are considering.}. 
However, we can still produce a warped 10D spacetime by placing a stack of $N$ D3-branes at the tip of the RC, extended along the 4 noncompact spacetime directions. The resulting 10D spacetime is called the warped resolved conifold\footnote{This is not to be confused with the similarly named \textit{resolved warped deformed conifolds} of \cite{Dymarsky2006,Butti2005}  based on the work of \cite{Papadopoulos:2000gj}. } (WRC) 
\cite{Zayas2000,Klebanov2007}.

In this paper we do not explicitly address the issue of closed string moduli stabilization. Instead, we assume that these are stabilized by some mechanism at a higher energy scale, decoupled from the open string brane moduli, which remain light. This allows us to investigate the inflationary dynamics from branes in the WRC. 

This assumption does not seem so strange when one remembers that the throat region is only an approximation to a fully compactified Calabi-Yau manifold. Multi-throat scenarios have been considered in which different throats are attached to different parts of the bulk Calabi-Yau \cite{Chen:2004gc,Cai:2008if,Cai:2009hw,Emery:2012sm,Emery:2013yua}. One of these other throats could support fluxes and or other mechanisms to stabilize the closed string moduli at a high energy scale. An additional warped throat may be necessary to embed the standard model, and another to uplift to a dS vacuum. With this multi-throat picture in mind, we assume an RC throat in which inflation occurs, and allow for other throats in which fluxes may be present which stabilize the complex structure moduli. In future work we hope to investigate closed string moduli stabilization in the WRC, and its effect on the brane potential.

In Section~\ref{sec:NI}, we model the inflaton as one of the angular coordinates of a probe wrapped D5-brane moving within the tip region of the WRC. This angular coordinate is periodic, and so has a shift symmetry, appearing in the brane potential inside a cosine. Once rescaled to have canonical kinetic terms, it appears with a decay constant, set by the choice of D-brane. In this way, we find a model of Natural Inflation from string theory. In taking a path in field space along the angular direction, we evade the Baumann-McAllister bound, which assumes the motion is in the radial direction
\cite{Baumann:2006cd}.

Finally, it should be noted that the work in this paper has a more general context beyond its application to the model of Natural Inflation presented in Section~\ref{sec:NI}. The authors of \cite{2006JHEP...11..031B, 2007PhRvL..99n1601B,2009JHEP...03..093B,2010JHEP...06..072B,Gandhi:2011id}
have developed a  systematic programme investigating the various corrections to the D-brane potential, corresponding to the potential on the Coulomb branch of the dual gauge theory, for perturbations of the Lagrangian. This includes various compactification effects, such as the deformation of the throat from gluing it to the bulk, the inclusion of IASD fluxes and a finite 4D curvature. The approach makes use of perturbative expansions around a Calabi-Yau background, requiring knowledge of the eigenfunctions of the Laplace operator on the unperturbed, unwarped Calabi-Yau background. Since the WDC has been extensively studied for moduli stabilization, the DC geometry was used to derive explicit expressions for the D-brane potential. 

However, the eigenfunctions of the Laplace operator on the DC are only known in the region well away from the tip of the DC. Interestingly, exact analytic solutions to the Laplace equation on the RC are known. In contrast to the DC case, these solutions are valid anywhere within the RC \cite{Klebanov2007}. Thus by studying the WRC we are able to provide a new explicit example of the general formalism developed in \cite{2006JHEP...11..031B} -   \cite{Gandhi:2011id}, which extends deeper into the IR. Thus, aside from inflationary applications, we believe studying D-brane dynamics in the WRC geometry is an interesting new avenue to explore from the perspective of the holographic dual gauge theory.
 
This paper is structured as follows. In Section~\ref{sec:sugra} we review the supergravity background arising from flux compactifications of 10D type IIB string theory, focussing on the various sources that can contribute to the D-brane potential. We also introduce the geometry of the WRC. In Section~\ref{sec:NI} we apply this to produce a Natural Inflation model from a D-brane in the WRC. We find that for the simple case of a D3-brane it's not possible to obtain a Planckian decay constant, as required by observational constraints. 
Indeed, in the WDC case, it was shown in \cite{Pajer:2008uy} that one can generate a Planckian decay constant from a large number of D3-branes, but the backreaction effects cannot then be ignored. 
By considering instead a wrapped D5-brane with flux we find that we can choose parameters such that the decay constant is Planckian, while the backreaction remains small. We conclude in Section~\ref{sec:conc} and provide scope for future research. We gather more technical aspects, such as estimations of the backreaction in Appendix~\ref{backreactionappend} and corrections to the potential coming from the 4D Ricci scalar in Appendix~\ref{ricciappend}. 

\section{The Supergravity Background} \label{sec:sugra}

\subsection{Flux Compactifications}
We would like to investigate the dynamics of D-branes in the WRC, which is a 10D warped geometry. Following the seminal work of \cite{Giddings2002}, these warped geometries arise naturally when local brane sources are present, and fluxes are non-trivial. Compactifications on such backgrounds are known as flux compactifications. 
We will briefly review the setup here to set our notation and introduce warped spacetimes. 

At leading order in $\alpha'$ and $g_s$, the type IIB action for bosonic fields, together with local sources, is given in the Einstein frame as
\begin{align}
\begin{split}
\tilde{S} = &-\frac{1}{2\kappa_{10}^{2}}\left [\int_{M_{10}}d^{10}X\sqrt{|g|}\left(R - \frac{|\p\tau|^2}{2(\text{Im}\tau)^2} - \frac{|G_3|^2}{2\text{Im}\tau} - \frac{|\tilde{F}_5|^2}{4\cdot5!} \right) 
+ \frac{1}{4i}\int_{M_{10}} \frac{C_4 \wedge G_3 \wedge \overline{G}_3}{\text{Im}\tau} \right ] 
\\
&+ S_{\text{loc}}\label{IIBaction}
\end{split}
\end{align}
where $S_{\text{loc}}$ is the action for the local sources. The form fields are defined as
\begin{align}
\tilde{F}_5 \equiv F_5 - \half C_2\wedge H_3 + \half B_2 \wedge  F_3,
\qquad
G_3 \equiv F_3 -\tau H_3
\end{align} 
with ${F}_5=dC_4$, ${F}_3=dC_2$, ${H}_3=dB_2$ and the axiodilaton is $\tau = C_0 + ie^{-\Phi}$, with $\Phi$ the dilaton. Here $\kappa_{10} = \frac{1}{2}(2\pi)^{4}g_sl_s^4$, with $l_s$ the string length. We must impose self duality of $\tilde{F}_5 = \star_{10}\tilde{F}_5$ by hand.

We assume a warped background metric ansatz
\begin{align}
ds^2 &= e^{2A(y)}g_{\mu\nu}dx^{\mu}dx^{\nu} +e^{-2A(y)}\tilde{g}_{mn}dy^{m}dy^{n}. \label{warped}
\end{align}
We have a maximally symmetric external 4D spacetime $X_4$, with metric $g_{\mu\nu}(x)$ and an internal unwarped space $\tilde{Y}_6$ with metric $\tilde{g}_{mn}(y)$. The warp factor is denoted $\mathcal{H}(y) \equiv e^{-4A(y)}$. 
We denote the warped internal 6D space without a tilde ${Y}_6$, with warped metric $g_{mn} = \mathcal{H}(y)^{1/2} \tilde{g}_{mn}$. Here, and in what follows, a tilde denotes use of the unwarped 6D metric $\tilde{g}_{mn}$ (except for $\tilde{F}_5$, which is just conventional notation in supergravity). 
Note that if the warp factor is shifted by a constant, this can be absorbed into a rescaling of the $x^{\mu}$ coordinates, so that only the functional dependence of $\mathcal{H}(y)$ is important.

For $\tilde{F}_5$ we take the ansatz 
\begin{align}
\tilde{F}_5 = (1 + \star_{10})  \ d\alpha(y) \wedge \sqrt{-\det g_{\mu\nu}} 
dx^0\wedge dx^1\wedge dx^2\wedge dx^3.\label{poinans}
\end{align}
Similarly to the warp factor, shifts of $\alpha(y)$ by a constant are irrelevant for $\tilde{F}_5$.

The inclusion of local sources, such as D-branes, leads to a non-trivial warp factor, producing a warped metric, together with non-vanishing fluxes. Varying the action leads to 
the Bianchi identity 
\begin{align}
d \tilde{F}_5 = H_3 \wedge F_3 + 2 \kappa_{10}^2 T_3 \star_6 \rho_3^{\text{loc}} \label{bian}
\end{align}
where $\star_6$ is the hodge dual in the warped 6D internal space with metric $g_{mn}$. Here $\rho_3^{\text{loc}}$ is the D3-brane charge density from the local sources.  Using  the warped spacetime \eqref{warped}, and the 5-form flux \eqref{poinans}, the Bianchi identity \eqref{bian} becomes
\begin{align}
\tilde{\nabla}^2  \alpha &= ie^{2A}\dfrac{G_{mnp}\star_6\overline{G}^{mnp}}{12\text{Im}\tau} + 2e^{-6A}\p_m \alpha \p^m e^{4A} + 2 \kappa_{10}^2e^{2A}T_3\rho_3^{\text{loc}} \label{bianchialphagen}
\end{align}
with ${\tilde{\nabla}}^{2} $ the Laplacian with respect to  the unwarped internal 6D metric. 
The trace of the Einstein equations can be written \cite{Giddings2002}, \cite{2010JHEP...06..072B},
\begin{align}
\begin{split}
\tilde{\nabla}^2  e^{4A} &=\mathcal{R}_4+  \frac{\kappa_{10}^2}{2}e^{2A}J^{\text{loc}} 
+ e^{2A}\dfrac{G_{mnp}\overline{G}^{mnp}}{12\text{Im}\tau} + e^{-6A}(\p_m\alpha\p^m\alpha + \p_m e^{4A}\p^m e^{4A})  \label{Eineee}
\end{split}
\end{align}
where 
\begin{align}
J^{\text{loc}} \equiv \frac{1}{4}\left (\sum_{M=4}^{9} {{T}_M}^M - \sum_{M=0}^{3} {{T}_M}^M \right )^{\text{loc}}.
\end{align}
Note that the local stress-energy tensor $T_{MN}^{\text{loc}}$ is contracted using the full 10D metric. 
The 4D Ricci scalar $\mathcal{R}_4$ is not present when the external spacetime is taken to be Minkowski \cite{Giddings2002}, but in the case of inflation, we take the external 4D spacetime to be quasi-de Sitter, with $\mathcal{R}_4 \approx 12H^2$ \cite{2010JHEP...06..072B}. 
We can combine
\eqref{bianchialphagen} and \eqref{Eineee} to give
\begin{align}
\begin{split}
{\tilde{\nabla}}^{2}  \Phi_- = \mathcal{R}_4 + \frac{e^{8A(y)}}{6\text{Im}\tau}|G_-|^2 + e^{-4A(y)}&|\partial \Phi_-|^2 + 2\kappa_{10}^2 e^{2A(y)}(J^{\text{loc}} - T_3\rho_3^{\text{loc}}) \label{fluxcom}
\\
\text{where } \Phi_- \equiv e^{4A(y)} - \alpha(y),  &\qquad  G_- \equiv \star_6 G_3 - iG_3. 
\end{split}
\end{align}
Note that $\Phi_-$ is insensitive to constant shifts. 

For the time being, we work in the noncompact warped volume limit for the internal geometry, $\text{vol}(Y_6) \to \infty$. But the warped volume is related to the reduced Planck mass $M_p$ by
\begin{align}
M_p^2 &= \frac{\text{vol}(Y_6)}{\kappa_{10}^{2}}
\end{align}
so that $M_p \to \infty$ in this noncompact limit. Then by the  Freidmann equation, $H^2 = V/(3M_p^2) \to 0$, so that the Ricci scalar for dS space vanishes $\mathcal{R}_4 \to 0$. We will later consider corrections from taking the compact limit with finite $M_p$.
In the noncompact limit, \eqref{fluxcom} becomes
\begin{align}
{\tilde{\nabla}}^{2}  \Phi_- = \frac{e^{8A(y)}}{6\text{Im}\tau}|G_-|^2 + e^{-4A(y)}&|\partial \Phi_-|^2 + 2\kappa_{10}^2 e^{2A(y)}(J^{\text{loc}} - T_3\rho_3^{\text{loc}}). \label{noncomphi}
\end{align}

Many well-understood local sources satisfy a BPS-like condition
$
J^{\text{loc}} \geq  T_3\rho_3^{\text{loc}}
$. D3-brane sources saturate this, while D5-branes satisfy but don't saturate it. Integrating \eqref{noncomphi} and assuming no boundary contribution at infinity, the LHS will vanish as it's a total derivative. 
But since each term on the RHS is positive semi-definite, each must individually vanish at leading order, giving an imaginary self-dual (ISD) solution 
$G_-=0$ and $\Phi_-=0$.

\subsection{The Warped Resolved Conifold}
We don't know the explicit metric on any smooth compact Calabi-Yau in 3 complex dimensions\footnote{Although progress has been made numerically, for example in \cite{Headrick:2005ch,Headrick:2009jz}.}. One way forward is to approximate a region of the fully compactified manifold  with a section of a noncompact throat, which satisfies the Ricci flat and K\"{a}hler condition, upon which we know the metric.  We then cut off this noncompact throat and glue it onto a bulk compact Calabi-Yau producing a fully compactified geometry. If we assume that the physics we're interested in is localised within the throat region then we have access to an explicitly known metric.  We can try to quantify the effects of the gluing procedure via perturbative expansions, assuming the gluing region is suitably far away.

The moduli space of Calabi-Yau threefolds has curvature singularities called conifold singularities. The space associated to the conifold singularity is known as \textit{the singular conifold} (SC)  \cite{Candelas1990}, which is a cone over the coset space $T^{1,1}$. On the plus side, it's Ricci flat and K\"{a}hler, and we know the explicit metric on it. However, it's noncompact and has a conical singularity at the tip, $r=0$. There are two topologically distinct ways of removing the conical singularity of the SC, while preserving the Ricci flat and K\"{a}hler conditions, arriving at \textit{the deformed conifold} (DC) \cite{2000JHEP...08..052K} and \textit{the resolved conifold} (RC) \cite{Zayas2000}, \cite{Klebanov2007}. For each of these, the metric is explicitly known. These are all noncompact, and must be truncated and glued to a bulk compact Calabi-Yau, as discussed in Section \ref{gluingsec}.

We often speak of a warped throat spacetime, which is a 10D warped spacetime \eqref{warped} involving a noncompact throat as the unwarped 6D space, $\tilde{Y}_6$, which can be approximated by a cone over some $X_5$ base in the large $r$ limit. These throats are a general family which include the SC, the DC and the RC. 

We will take the internal unwarped 6D manifold to be the RC, using the coordinates $y^m = (r,\psi,\theta_1,\phi_1,\theta_2,\phi_2)$ in which the metric takes the form  \cite{Zayas2000}
\begin{align}
\begin{split}
ds_{\text{RC}}^2  = \tilde{g}_{mn} dy^m dy^n = &\kappa^{-1}(r)dr^2 + \frac{1}{9}\kappa(r)r^2(d\psi + \cos \theta_1 d\phi_1 +  \cos \theta_2 d\phi_2)^2 
\\
& + \frac{1}{6}r^2(d\theta_1^2 + \sin^2 \theta_1 d\phi_1^2) + \frac{1}{6}(r^2 + 6u^2)(d\theta_2^2 + \sin^2 \theta_2 d\phi_2^2), 
\end{split} \label{resolvedcon}
\\ \nonumber
\end{align}
\text{with }
\begin{align}
\kappa(r) &= \frac{r^2 + 9u^2}{r^2 + 6u^2}.
\end{align}
Here $u$ is called the \textit{resolution parameter}, which has dimensions of length. This naturally defines a dimensionless radial coordinate $\rho \equiv r/(3u)$. As $r \to 0$ the second and third parts of \eqref{resolvedcon} vanish, corresponding to a shrinking $S^3$ with coordinates $(\psi,\theta_1,\phi_1)$, leaving an $S^2$ with coordinates $(\theta_2,\phi_2)$ of radius $u$.
 
One can consider the 10D geometry sourced by placing a stack of $N$ D3-branes extended along the noncompact 4 spacetime dimensions, appearing pointlike localized at the north pole, $\theta_2=0$, of the $S^2$ at the tip of the RC \cite{Klebanov2007}. The resulting geometry is a warped spacetime called the warped resolved conifold (WRC), with 10D metric  \cite{Zayas2000}, \cite{Klebanov2007}
\begin{align}
ds^2  =\mathcal{H}^{-1/2}(\rho,\theta_2)ds_{\text{FRW}}^2 + \mathcal{H}^{1/2}(\rho,\theta_2)ds^2_{RC}
\end{align}
where we have taken the 4D spacetime to be FRW, for our cosmological application, and the 6D unwarped space is the RC. 

The warp factor, $\mathcal{H}(\rho,\theta_2)$, is the solution to the Green's function equation for the Laplace operator on the RC. An exact expression for the WRC warp factor is  \cite{Klebanov2007} 
\begin{align}
\label{warp}
\mathcal{H}(\rho, \theta_2) &=  (L_{T^{1,1}}/3u)^4 \sum_{l=0}^{\infty}  (2 l+1) H^A_l (\rho) P_l[cos(\theta_2) ],
\end{align}
with the $T^{1,1}$ lengthscale set by $L_{T^{1,1}}^4 = (27/4) \pi N  g_s l_s^4$. The $P_l$ are the Legendre polynomials, and the radial functions  $H^A_l(\rho) $ are given in terms of the $_2F_1(a,b,c;z) $ hypergeoemetric functions as 
\begin{align}
H^A_l (\rho) =  \frac{2{\tilde C}_\beta}{\rho^{2+2\beta}}\, {_2F_1}(\beta, 1+\beta, 1+2 \beta; -1/\rho^2 ) \label{Ha}
\\
\text{where }{\tilde C}_\beta = \frac{\Gamma(1+\beta)^2 }{\Gamma(1+2 \beta) } , \qquad \beta = \sqrt{1 +(3/2)l(l+1)}.
\end{align}
Since localizing the stack at the north pole specifies an angle, the warp factor must now have both angular and radial dependence - whereas the warp factors only depend on the radial coordinate in the case where the internal geometry is the SC or the DC, and is an assumption usually made for generic warped throats. This motivates us to explore branes moving in the angular directions within the WRC, where the warping also acts in the angular direction.

The gauge gravity correspondence for a stack of D-branes near a conical singularity  of a cone over an $X_5$ base was investigated in generality in \cite{1998NuPhB.536..199K}  where strings on $AdS_5 \times X_5$  correspond to a certain dual $\mathcal{N}=1$ gauge theory. Theories where the stack of D-branes are localized at a point on the resolution of an orbifold have also been considered, for example in \cite{Krishnan2008a} for $\mathbb{C}^3/\mathbb{Z}_3$.

The WRC has been investigated from the point of view of the gauge gravity correspondence in \cite{Klebanov2007}, which found the dual gauge theory living on the stack to be a  4D $\mathcal{N}=1$, $SU(N) \times SU(N)$ gauge theory coupled to two chiral superfields ${A}_i, i=1,2$ in the $(\textbf{N},\overline{\textbf{N}})$ representation of $SU(N) \times SU(N)$; and two chiral superfields ${B}_j, j=1,2$ in the $(\overline{\textbf{N}}, \textbf{N})$ representation of $SU(N) \times SU(N)$. The fields ${A}_i, {B}_j$ are given VEVs such that the operator 
\begin{align}
\mathcal{U} \equiv \frac{1}{N}Tr( | {B}_1|^2 + | {B}_2|^2 - |{A}_1|^2 - |{A}_2|^2)
\end{align} 
has VEV $\braket{\mathcal{U}} = u^2$. The moduli space of these vacua has exactly the geometry of the RC, with  resolution parameter $u$.

\subsection{Gluing a Warped Throat to a Bulk Calabi-Yau} \label{gluingsec}

For model building purposes we first take the noncompact limit, with an infinitely long warped throat. This gives an ISD flux solution, with $G_-= 0=\Phi_-$. 
However, note that a 10D geometry with an infinitely long warped throat does not lead to 4D dynamical gravity, as $M_p$ is infinite.

To remedy this, we cut off the warped throat at some large radial distance $r_{UV}$, and glue it to a compact bulk Calabi-Yau. 
The metric on the bulk is not known, but the metric on the warped throat is explicitly known for certain warped throats, such as the SC, the DC and the RC. For this reason we try to get inflation to occur within the warped throat. Although this may not be generic,  it allows us to do calculations. Perturbations of $\Phi_-$ arise as a result of this gluing procedure and are solutions to the Poisson equation \eqref{fluxcom}, written in terms of the warp factor as
\begin{align}
{\tilde{\nabla}}^{2} \Phi_- &= \mathcal{R}_4 + \frac{\mathcal{H}^{-2}}{6\text{Im}\tau}|G_-|^2 + \mathcal{H}|\partial \Phi_-|^2 + 2\kappa_{10}^2 \mathcal{H}^{-1/2}(J^{\text{loc}} - T_3\rho_3^{\text{loc}}).  \label{mastereq}
\end{align}
We will assume that the gluing will induce corrections of $\mathcal{O}(\delta)$ to $\Phi_-$, for some small $\delta$, and we assume corrections to $G_-$ are of the same order. We will assume also that the local terms don't contribute at this order. The $G_-$ and $\Phi_-$ terms appear on the RHS of \eqref{mastereq} at second order in $\delta$, so that the leading order perturbation of $\Phi_-$ in the large throat limit is a solution to the homogeneous Laplace equation, so we denote it with subscript $h$,
\begin{align}
{\tilde{\nabla}}^{2} \Phi_h &= 0 .\label{laplacephi-}
\end{align}
 
This distinguishes it from $\Phi_-$ which is the full solution to the Poisson equation arising when we consider the effect of a non-negligible $\mathcal{R}_4$, where the leading order correction to $\Phi_- = 0$ from the gluing is given by 
\begin{align}
{\tilde{\nabla}}^{2} \Phi_- &= \mathcal{R}_4 \label{poissonphi-}
\end{align}
so that  $\Phi_-$ includes $\Phi_h$, but also the particular solutions to the Poisson equation.

The solutions of \eqref{laplacephi-} and \eqref{poissonphi-} will depend on the unwarped internal 6D geometry. In subsubsections \ref{gluedef} and \ref{glueres}  we consider solutions to the Laplace equation \eqref{laplacephi-} for the DC and the RC geometry respectively. In Appendix~\ref{ricciappend} we consider solutions to the Poisson equation \eqref{poissonphi-} for the RC.

\subsubsection{Gluing the Deformed Conifold}  \label{gluedef}

Unfortunately, the exact solutions to the Laplace equation \eqref{laplacephi-} are not known for the deformed conifold. Progress has only been made in the mid-throat region \cite{2009JHEP...03..093B} , $r_{\text{IR}} \ll r \ll r_{UV}$, where the geometry approximates that of the singular conifold, with the 10D metric approaching $AdS_5 \times T^{1,1}$. 

We can expand the solution in terms of the eigenfunctions $Y_{L}(Z_i)$ of  the 5D Laplacian on $T^{1,1} $ as
\begin{align}
\Phi_h(y) = \sum_L \Phi_L(r)Y_{L}(Z_i)
\end{align}
where $Z_i $ are the angular coordinates  on $T^{1,1}$. Here, the multi-index $L\equiv (l_{1},m_{1},l_{2},m_{2},R)$, labels  the $SU(2)_1\times SU(2)_2\times U(1)_{R}$ quantum numbers under the corresponding isometries of $T^{1,1}$.

But the equation for the radial part, $\Phi_{L}(r)$, has no known analytic solution for the DC, and can only be solved numerically \cite{Krishnan2008}. Limited to the mid-throat region, $\Phi_h$ can be expanded in powers of $ r/r_{\rm UV}$ as \cite{Gandhi:2011id}
\begin{align}
\Phi_{h}(r,Z_i) &\approx \sum_{L}c_{L}
\left (\frac{r}{r_{\rm UV}}\right )^{\Delta(L)}Y_{L}(Z_i) \label{modes}
\\
\text{where } \Delta(L) &\equiv -2+\sqrt{6[l_{1}(l_{1}+1)+l_{2}(l_{2}+1)-{R^2}/8]+4}
\end{align}
where $c_{L}$ are constant coefficients. 

The lowest value of $\Delta(L)$ will give the leading contributions for $r<r_{\rm UV}$. The lowest value is $\Delta(L)=3/2$, for $L=(1/2,1/2,1/2,1/2,1)$ \cite{2009JHEP...03..093B}. 
But the $U(1)_R$ symmetry of $T^{1,1}$ is broken in the DC to a discrete $\mathbb{Z}_2$, so only modes with $R=0$ are allowed, forbidding the $\Delta(L)=3/2$ mode.
The next smallest mode, $\Delta(L) = 2$, for $L=(1,0,0,0,0)$ or $L=(0,0,1,0,0)$ is allowed, and comes with an angular term $Y_L \sim \cos \theta_i$. This mode was analyzed in \cite{Gregory:2011cd} in the DBI limit.

The coefficients  $c_{L}$ appearing in \eqref{modes} are undetermined, apart from their small size. The authors of 
\cite{Agarwal:2011wm}, \cite{2012JCAP...06..020D} have taken a statistical approach to investigating warped D3-brane inflation in this approximation. For example,   \cite{2012JCAP...06..020D} explores the parameter space spanned by the first 12 $c_{L}$ coefficients and determining the success/failure of the model in each case. However, a flat field space metric was used instead of a curved conifold metric which was corrected for in \cite{2012JCAP...10..046M} and appeared in a corrected version of \cite{2012JCAP...06..020D}. This statistical approach has been restricted to the mid-throat region of the singular conifold, but could also be applied to the RC geometry we consider in this paper, with the benefit of not being restricted to lie in the mid-throat region.

\subsubsection{Gluing the Resolved Conifold} \label{glueres}
For the RC we can again expand $\Phi_h$ in the $Y_L(Z_i)$, but in this case the radial part of the Laplace equation can be solved exactly on the resolved conifold in terms of hypergeometric functions \cite{Klebanov2007}, which was not the case for the DC.

In this paper, we are interested in probing the tip of the WRC, so we focus on solutions of the Laplace equation which are invariant under the $SU(2) _1\times U(1)_{\psi}$ which rotates the $(\theta_1,\phi_1)$ and $\psi$ coordinates of the shrinking $S^3$ which has zero radius at the tip\footnote{It should be emphasised there is no particular reason other than simplicity in focussing on solutions with the given isometries.
We focus on dynamics at the tip, as this is new analytic territory compared to that available for the WDC. 
 Note that studying  dynamics along the whole throat is calculable in the WRC but we leave this interesting problem to a future investigation.}. This leaves us $(\rho,\theta_2,\phi_2)$, from which we now drop the subscripts. 

There are two particular independent solutions to the radial part of the Laplace equation on the RC, invariant under the $SU(2)_1\times U(1)_{\psi} $. They are $H^A_l(\rho)$, given in \eqref{Ha}, and $H^B_l(\rho)$, given by
\begin{align}
H^B_l (\rho) = {_2F_1}(1-\beta, 1+\beta, 2; - \rho^2 ).
\end{align}

The most general solution to the Laplace equation with the given isometries is 
\begin{align}
\Phi_{h}(\rho, \theta, \phi) = \sum_{l=0}^\infty\sum_{m=-l}^{m=l} [a_l H^A_l(\rho) +b_l H^B_l(\rho) ]Y_{lm}( \theta,\phi). \label{potwrc}
\end{align}

This solution is valid \textit{anywhere} within the WRC throat, in particular near the tip. This is a much better situation compared to the WDC, where the solution is only valid in the mid-throat region. The coefficients are undetermined, yet small, $a_l, b_l = \mathcal{O}(\delta)$. 

For completeness, we give the asymptotics of the two radial functions $ H^A_l(\rho)$ and $ H^B_l(\rho)$ \cite{Klebanov2007}
\begin{align}
\frac{2}{ \rho^2} + 4 \beta^2  \log \rho  + {\cal O}(1)\quad  \overset{0 \leftarrow \rho}{\longleftarrow} \quad  &H^A_l(\rho)  \quad  \overset{\rho \rightarrow \infty}{\longrightarrow} \quad
  \frac{2 {\tilde C}_\beta}{ \rho^{2+2\beta} } 
\\ \nonumber
 {\cal O}(1)\quad  \overset{0 \leftarrow \rho}{\longleftarrow} \quad  &H^B_l(\rho)  \quad  \overset{ \rho \rightarrow \infty}{\longrightarrow} \quad {\cal O}( \rho^{-2+2\beta } ). 
\end{align}

In Appendix~\ref{ricciappend} we will consider solutions to the Poisson equation \eqref{poissonphi-} on the RC, which gives corrections to $\Phi_-$ in the limit of a finite but large throat.

\subsection{Moduli Stabilization}

In an ISD flux compactification, the complex-structure moduli and the axiodilaton $\tau$ experience a potential, coming from the following term in the 10D type IIB action \eqref{IIBaction}
\begin{align}
V_{\text{flux}} &= \frac{1}{2\kappa_{10}^{2}}\int d^{10}X\sqrt{|g|}\left [ - \frac{|G_3|^2}{2\text{Im}\tau} \right ], \qquad |G_3|^2 \equiv \frac{1}{3!}g^{MN}g^{PQ}g^{RS}G_{MPR}G^*_{NQS}. \label{g3flux}
\end{align}

The 4D effective description of an ISD flux compactification can also be derived from a K\"{a}hler potential and a Gukov-Vafa-Witten flux superpotential, $W$, of $\mathcal{N}=1$ supergravity \cite{2000NuPhB.584...69G}
\begin{align}
W = \int_{\tilde{Y_6}} \Omega \wedge G_3 \label{superpot}
\end{align}  
where 
$\Omega$ is the holomorphic 3-form associated to the Calabi-Yau $\tilde{Y}_6$. 
$\mathcal{N}=1$ supersymmetry is preserved if $G_3$ is a primitive $(2,1)$-form \cite{Giddings2002}. 
Since $G_3$ depends on the axiodilaton and $\Omega$ depends on the complex-structure moduli, \eqref{superpot} is independent of the K\"{a}hler moduli. Since the  K\"{a}hler potential is of no-scale type, the resulting scalar potential stabilizes only the complex-structure moduli and axiodilaton. The K\"{a}hler moduli are assumed lighter than the complex-structure moduli and axiodilaton, and so once these heavier moduli have been integrated out, the K\"{a}hler moduli can then be stabilized by quantum non-perturbative effects, such as wrapped D7-branes. This leads to an AdS vacuum, which requires uplifting to a dS vacuum involving the inclusion of anti-brane sources \cite{Kachru:2003sx}. 

Previous studies \cite{Giddings2002}, \cite{Kachru:2003aw}, \cite{Kachru:2003sx} have stabilized the complex-structure moduli and axiodilaton using the DC as the internal 6D unwarped metric. 
The DC has Hodge numbers $h^{2,1}=1, h^{1,1}=0$ \cite{Aganagic:1999fe} so there is only one complex-structure modulus to stabilize -- the deformation parameter of the deformed conifold, $z$. Since $h^{2,1}=1$, one can turn on primitive $(2,1)$-form fluxes which preserve $\mathcal{N}=1$ SUSY.
The third betti number is $b_3 = 2 + 2h^{2,1} =4$, so there are 4 non-trivial 3-cycles in the DC.
One of these, the $A$-cycle, is associated to the finite size $S^3$ at the DC tip. There is an associated 3-cycle, $B$, which intersects this $A$-cycle exactly once.

One can then choose to turn on the following quantized $(2,1)$-form fluxes through the $A, B$ cycles
\begin{align}
\frac{1}{2\pi\alpha'}\int_A F_3 = 2\pi M \qquad \text{and} \qquad \frac{1}{2\pi\alpha'}\int_B H_3 = - 2\pi K. \label{fluxquant}
\end{align}
These fluxes allow for the superpotential to be written in terms of $z$. Since SUSY is preserved, one can minimize the scalar potential by imposing $D_zW=0$ which stabilizes $z$, by solving $D_zW =0$ for $z$. In the noncompact DC, the axiodilaton is not fixed by the superpotential - instead it is frozen in the Klebanov-Strassler solution.

In the noncompact DC limit, with an infinitely long throat, the DC $B$-cycle degenerates to infinite size. When the DC is cut off and glued to a compact bulk Calabi-Yau, the $B$-cycle becomes finite. The gluing will generically increase $h^{2,1}$ for the entire manifold, meaning there are more complex-structure moduli to stabilize. Assuming one can first stabilize $z$ near the conifold point $z=0$, the additional complex-structure moduli can be stabilized while preserving $\mathcal{N}=1$ supersymmetry using the superpotential generated by the fluxes. In this compact case, the axiodilaton is no longer frozen, as in the KS solution. It is now fixed by including $(2,1)$-form fluxes over the remaining two 3-cycles distinct from $A$ and $B$. This contributes to $W$, and one can then impose $D_{\tau}W =0$, near $z=0$ and solve for $\tau$, which preserves $\mathcal{N}=1$ SUSY \cite{Giddings2002}.

The RC on the other hand has Hodge numbers $h^{2,1}=0, h^{1,1}=1$ \cite{Aganagic:1999fe} so there are no complex-structure moduli to stabilize, instead there is a single K\"{a}hler modulus, the resolution parameter $u$. Since $h^{2,1}=0$, there are no cohomologically nontrivial closed $(2,1)$-forms, so one can't turn on fluxes which preserve $\mathcal{N}=1$ SUSY.
The third betti number is $b_3 = 2 + 2h^{2,1} =2$, so there are 2 non-trivial 3-cycles in the RC, on which $(3,0)$-form fluxes could be turned on,  and these would classically fix the axiodilaton. However, these fluxes will break $\mathcal{N}=1$ SUSY, and so the equation determining the vev of the axiodilaton would be obtained by minimizing the full scalar potential, rather than just solving $D_{\tau}W =0$. It would be interesting to compare the energy scale at which SUSY would need to be broken to fix the axiodilaton in this way, compared to the energy scale at which SUSY is broken when uplifting to a dS mimimum, as in \cite{Kachru:2003aw}.

The K\"{a}hler modulus $u$ would need to be stabilized by non-perturbative effects, \cite{Kachru:2003sx}. This may also fix the axiodilaton, without the need for breaking SUSY. It would be interesting to investigate at what value $u$ is fixed at, and its possible mixing with the open string brane moduli.

When the compact case is considered, the bulk Calabi-Yau may have a different topology, allowing for $h^{2,1}>0$, and so SUSY preserving primitive $(2,1)$-form fluxes may be turned on, stabilizing the additional complex-structure moduli and the axiodilaton. 

Indeed, there have been multi-throat scenarios proposed, \cite{Chen:2004gc}, in which the standard model is required to be situated in a seperate throat to that where inflation occurs, and to where the anti-branes, which end inflation, are located. Since this scenario allows for these extra throats, it doesn't seem too much to ask that there is another warped throat in the compactification, perhaps a WDC throat, which allows stabilization of the complex-structure moduli. Perhaps it's too much to hope that one throat will be able to do everything: stabilize, inflate, produce the standard model and give a dS vacuum. 

In addition, it has been suggested that there might be a mild hierarchy between the scales at which the closed string moduli and open string moduli are stabilized, so that the two problems are approximately decoupled \cite{Marchesano2014}. This serves as motivation for us to investigate brane inflation in the WRC.

\section{Natural Inflation Model} \label{sec:NI}
\subsection{Preview}

We are interested in modelling inflation using the open string moduli arising as the coordinates of a probe D-brane within the WRC geometry, which are scalar fields of the 4D effective theory. The probe approximation means ignoring any backreaction coming from the mobile brane, onto the WRC supergravity background, which is itself sourced by the stack of $N$ D3-branes at the north pole of the $S^2$ at the tip of the RC. Note that $N$ must be large for the SUGRA solution to be valid, so that for a single probe D3-brane, the backreaction should be negligable. The backreaction for a probe D5-brane is discussed in Appendix~\ref{backreactionappend}.

In the original warped throat models of inflation, motion in the angular directions is assumed stabilized before inflation begins, with inflation occurring along a radial path.  But the Baumann-McAllister (BM) bound implies a stringent upper bound on the scalar-to-tensor ratio $r$ for motion along the radial path \cite{Baumann:2006cd}. In this work we instead look to inflate along the angular direction $\theta_2$, where the BM bound no longer restricts $r$, as pointed out in \cite{Langlois:2009ej}.

Indeed, contrary to the WSC and WDC, the simplest form of the warp factor for the WRC has dependence on both $\rho$ and $\theta_2$, so for the WRC one might expect interesting motion in the $\rho$ and $\theta_2$ directions. More general models can be considered where the brane also moves in the other directions - we defer this more complicated study to future investigation. 

Brane motion in both the radial and one angular direction of the WDC was considered in \cite{2007JHEP...04..026E,Easson:2007dh,Gregory:2011cd,Kaviani:2012qw,Kidani2014}, however, the majority of the trajectory was in the radial direction. This was done in the DBI limit, where the brane motion is ultrarelativstic.
Studies of brane inflation with multiple fields were considered more generally in \cite{Langlois:2008wt,Langlois:2008qf,Langlois:2009ej}. 

In this work, we will instead show that for a suitable choice of coefficients in the homogeneous solution $\Phi_h$ in \eqref{potwrc}, $\rho$  rapidly approaches  a minimum value near the tip of the WRC. The brane then follows a path in the angular direction $\theta_2$ down to the minimum of the potential\footnote{We shall see that with this choice of coefficients, the potential is flat along the other 4 angular directions and so we can choose to set these anglular fields  to zero. }. It is the latter motion which will generate the 50-60 e-folds of inflation. We can set the initial conditions to be such that we begin at the minimum in the radial direction and just off to the side of the maximum of the potential in the $\theta_2$ direction, which we now relabel as $\theta_2\equiv \theta$, without ambiguity.

Using this setup we will construct an explicit original model of \textit{Natural Inflation} \cite{Freese1990,PhysRevD.47.426,Freese:2004un,Freese2014} i.e. an inflationary model for the inflaton $\sigma$, with the potential 
\begin{align}
V(\sigma) = \Lambda^4\left [1  + \cos\left (\frac{\sigma }{f}\right )\right ]. \label{natpot1}
\end{align}
The observationally favoured values for the number of e-foldings $N$, the scalar spectral tilt $n_s$, together with an observably large tensor-to-scalar ratio $r$ \cite{Freese2014} are given by the parameter choice of energy scale $\Lambda = M_{\text{GUT}} \approx10^{16}$GeV  and decay constant $f \sim m_p \sim 5M_{p} \approx 1.2 \times 10^{19}$GeV, for $m_p$ the Planck mass, and $M_p$ the reduced Planck Mass. 

We will derive a potential of the form \eqref{natpot1} by considering brane potentials of the general form
\begin{align}
V &=\frac{M_p^2}{4} \Big \{ \varphi(y) + \lambda \left [ \overline{\Phi}_-(y) + \Phi_h(y)\right ] \Big \}\label{totalpot}
\end{align}
where $\varphi(y)$, $\lambda$, $\overline{\Phi}_-(y)$ and $\Phi_h(y)$ depend on the choice of probe D-brane. The term $\varphi(y)$ arises from the noncancellation of DBI and CS terms in the slow roll limit, and also includes any constants independent of the brane position which contribute to the 4D energy density. The $\Phi_h(y)$ term is the nonconstant solution to the homogeneous Laplace equation on the RC. The $\overline{\Phi}_-(y)$ term is the inhomogeneous part of the solution to the Poisson equation, present only when considering corrections from the Ricci scalar. This is explained in more detail in Appendix~\ref{ricciappend}. The factor of $\lambda$ is a constant which depends on the choice of probe D-brane.

The $\Phi_h(y)$ term is independent of the choice of probe brane, and one can freely choose the coefficients of independent solutions of the Laplace equation.  We choose to keep two independent solutions to the Laplace equation, but this is by no means a unique choice. It is motivated only by our aim of reaching a potential related to the Natural Inflation potential. 

One solution we keep is non-normalizable for large $\rho$, with charges 
\begin{align}
L=(l_1,m_1,l_2,m_2,R)=(0,0,1,0,0)
\end{align} which is present for the choice of non-zero $b_1$, and takes the form
\begin{align}
H^{B}_{1} = \frac{3}{2}(3\rho^2 + 2)\cos \theta.
\end{align}
This term is desirable because of its cosine term.
Our model will take the inflaton field $\sigma$ to be the canonical scalar field $\Theta$, proportional to the angular coordinate $\theta$. The normalisation of $\Theta$ in terms of $\theta$ will determine $f$ in \eqref{natpot1}, and depends on the choice of probe D-brane. 

We will also keep one mode which is normalizable for large $\rho$, with charges  
\begin{align}
L=(l_1,m_1,l_2,m_2,R)=(0,0,0,0,0)
\end{align} which is present for nonzero $a_0$, and takes the form
\begin{align}
H^A_0 = \frac{1}{\rho^2} - \log \left (\frac{1}{\rho^2} + 1\right ).
\end{align}
Taking $a_0>0$ gives a large positive contribution near $\rho  =0$. 
Thus, our choice of coefficients leads to the homogeneous solution to the Laplace equation on the RC, $\Phi_h$, given by
\begin{align}
\Phi_h
&= \frac{{a_0}}{\rho ^2}-{a_0} \log \left(\frac{1}{\rho ^2}+1\right)+\frac{3}{2} {b_1} \left(3 \rho ^2+2\right) \cos \left (\frac{\Theta }{5M_p}\right ). \label{homd5}
\end{align}

In Subsection~\ref{d3subsection}, we consider a probe D3-brane. This has $\lambda = 4T_3/M_p^2$, and a constant $\varphi(\rho)=V_0$. This is because the DBI and CS terms exactly cancel in the slow roll limit for a D3, but other sources contribute to the 4D energy density to give the constant $V_0$. This constant sources a mass term,  $\overline{\Phi}_- \sim m^2 \rho^2$ at leading order in small $\rho$, when one includes the $\mathcal{R}_4$ contribution \cite{2010JHEP...06..072B}. This mass term, when combined with the large positive wall from the $H^A_0$ term will give a radial minimum at small $\rho$. 

In Subsection~\ref{d5subsection}, we consider a probe wrapped D5-brane with electric flux turned on in the wrapped directions. In this case, $\varphi(\rho)$ is not constant, and  depends on $\rho$ quadratically, $\varphi(\rho) \sim \rho^2$. This arises from the non-cancellation of the DBI and CS terms in the slow roll expansion of the action. This $\rho^2$ term will again give a radial minimum at small $\rho$ when combined with the large positive contribution for small $\rho$ from the $H^A_0$ term. In Appendix~\ref{ricciappend} we show that $\varphi(\rho) \sim \rho^2$ sources a quartic term in $\overline{\Phi}_-$, with a large positive coefficient. However, this quartic term is subleading in the small $\rho$ limit. 

Our work differs from previous works deriving Natural Inflation from closed string axions in string theory, which arise from integration of a $p$-form over a $p$-cycle in the compact space. 
For these closed string axions, a periodic potential can arise in the 4D effective theory when the continuous axion shift symmetry is spontaneously broken to a discrete shift symmetry, due to nonperturbative effects arising from worldsheet instantons or Euclidean D-brane instantons.
The decay constant for each closed string axion is set by the kinetic terms, which depend on the type of axion used.
However, these decay constants turn out to be generically sub-Planckian, by looking at the kinetic terms for these axions and relating them to the compactification volume, and hence the Planck mass, together with the validity of the $\alpha'$ expansion \cite{McAllister:2008hb,Banks:2003sx,Svrcek:2006yi}.

It should be noted that although one can't generically obtain a Planckian decay constant for a single closed string axion from a controlled string theory, there may be additional structure which allows one to obtain a Planckian effective decay constant. For example, in \textit{Aligned Natural Inflation} \cite{Kappl2014a,Kim:2004rp}, the extra structure involves two interacting closed string axions with sub-Planckian decay constants, which after a suitable amount of fine tuning aligns to produce a direction in axion field space which has a Planckian effective decay constant. Further extensions to more than two axions were investigated in \cite{Choi2014}. Embedding axion alignment in string theory was recently explored in \cite{Long:2014dta}, using gaugino condensation on magnetized or multiply-wound D7-branes, however closed string moduli stabilization was not addressed in this model. In \cite{Ben-Dayan:2014lca}, moduli stabilization was included in both KKLT and LVS regimes, with non perturbative effects in the superpotential used to produce the alignment, or the alternative \textit{Hierarchical Axion Inflation} \cite{Ben-Dayan:2014zsa}. Another embedding of Aligned Natural Inflation in IIB orientifolds was discussed in \cite{Gao:2014uha}, using $C_0$ and $C_2$ R-R axions in the LVS regime.

A separate model, \textit{N-flation} \cite{Dimopoulos:2005ac},  motivated by \textit{Assisted Inflation} \cite{Liddle:1998jc}, and similar proposals \cite{Ashoorioon:2009wa,Ashoorioon:2011ki,Ashoorioon:2014jja}, use many light non-interacting closed string axions which contribute to one effective axion direction with a Planckian effective decay constant. 
The masses of the axions can be made hierarchically lighter than the K\"{a}hler moduli \cite{Cicoli:2014sva}. However, the large number of axions requried \cite{Kim:2006ys} can renormalize the Planck mass, spoiling the achievement of a Planckian decay constant. N-flation and Aligned Natural Inflation can be combined as in \cite{Bachlechner:2014hsa,Choi:2014rja,Czerny:2014qqa}. 

\textit{Axion Monodromy Inflation} encompasses a related class of models which achieve large field inflation from closed string axions in specific brane backgrounds from string theory and F-theory. The brane backgrounds explicitly break the axionic shift symmetry of the potential, giving rise to monodromy \cite{Silverstein:2008sg,2008JHEP...10..110B,McAllister:2008hb,Peiris:2013opa,Franco2014,Flauger:2009ab,Blumenhagen2014}. Inflation can continue through many cycles of the axion field space, with an effective field range which is much larger than the original period of the axion, giving large field inflation. However, the axion monodromy models generically suffer from a large backreaction problem, coming from a large D3-brane charge and/or backreaction from the branes themselves \cite{Conlon:2011qp}. These backreaction difficulties are alleviated when the monodromy mechanism is combined with the idea of alignment, in a model known as \textit{Dante's Inferno}, where the inflaton takes a gradual spiral path in 2D axion field space \cite{Berg:2009tg}.
Alternative monodromy models use D7-brane position moduli with a shift symmetry broken by a flux superpotential \cite{Hebecker2014}, or the Ignoble Approach of \cite{Kaloper:2011jz} where the axion mixes with a topological 4-form field strength to produce the monodromy.
Possible embeddings of Natural Inflation in supergravity were investigated recently in \cite{Kallosh2014}. A review of axion inflation in the Planck era is given in \cite{2013CQGra..30u4002P}.

Our Natural Inflation model doesn't use any of the above closed string axions as the inflaton, and so a Planckian decay constant is not ruled out \textit{a priori}. Our inflaton is instead an open string modulus, identified with the position of the brane in an angular direction on the RC. It has a discrete shift symmetry, which is set by the internal geometry, with the decay constant set by the normalization of the kinetic term. Thus, our model shares more similarity to models of spinflation \cite {Gregory:2011cd}. However, we also differ from the setup of \cite{Gregory:2011cd}, because we explore the use of the RC rather than the DC geometry. Also, we use a probe wrapped D5-brane with flux, rather than a probe D3-brane without flux, and we probe the tip rather than the mid-throat region. The combination of these choices allows us to select a Planckian decay constant, which we then check is suitable for a controlled supergravity approximation, and doesn't produce a large backreaction.

In previous work, inflation from a brane moving in an angular direction was found to be rather ineffective \cite {Easson:2007dh,Gregory:2011cd,Agarwal:2011wm}. In these models the initial conditions were such that the brane starts far from the tip and where the major contribution to the number of e-folds was from the radial motion towards the tip. However, in our model we make a different choice of initial conditions, such that the brane begins at the tip and due to the steep potential in the radial direction, experiences no motion in the radial direction. All of the inflationary e-folds occur along the angular direction. The flatness in the angular direction is a result of the choices of initial radial position (bringing a large warp factor), the use of a wrapped D5-brane (rather than  a D3-brane) and turning on 2-form flux through the wrapped dimensions of the brane (allowing for a nonzero CS potential from the gluing to a bulk Calabi-Yau).

%

\subsection{D3-brane in the WRC} \label{d3subsection}

To begin, we review how the potential, $V$, felt by a single slowly moving probe D3-brane in a warped throat geometry is  related to $\Phi_-$ via $V= V_0 + T_3\Phi_-$ \cite{2006JHEP...11..031B}. 

Consider a probe D3-brane, with worlvolume coordinates $\chi^{a}$ extended along the four noncompact directions, $M_4$. The action has contributions from the DBI term and the Chern-Simons term
\begin{align}
S_{D3} &= - T_3\int_{{M}_4}d^4\chi\sqrt{-\det(P_4[g_{MN} + B_{MN} + 2\pi\alpha'{F}_{MN} ])} + T_3\int_{{M}_4}P_4[C_4]
\end{align}
where $P_4$ is the pullback of the brane worldvolume to $M_4$. We assume 4D isotropy and homogeneity, relevant for cosmological spacetimes, meaning we should consider time-dependent internal coordinates $y^m(t)$. Substituting a general warped 10D metric $g_{MN}$ of the form \eqref{warped}
\begin{align}
ds^2 &= \mathcal{H}^{-1/2}(y)g_{\mu\nu}^{\text{FRW}}dx^{\mu}dx^{\nu} +\mathcal{H}^{1/2}(y)\tilde{g}_{mn}dy^{m}dy^{n} 
\end{align}
together with the ansatz \eqref{poinans} for $C_4$ and taking $B_2 = 0 = F_2$ gives the effective 4D Lagrangian density
\begin{align}
\mathcal{L} &= -T_3\mathcal{H}^{-1}(y)\sqrt{1-\mathcal{H}(y)\tilde{g}_{mn}\dot{y}^m\dot{y}^n}+T_3\alpha(y). \label{lagrangianwarp}
\end{align}

For slowly rolling fields, we can expand the square root in \eqref{lagrangianwarp}, to give
\begin{align}
\mathcal{L} &\approx \frac{1}{2} T_3 \,  \tilde{g}_{mn}\dot{y}^m\dot{y}^n -V(y) \label{lagrangianwarpexpand}
\end{align}
where the D3-brane potential $V(y)$ will be
\begin{align}
V(y) &= V_0 +  T_3(\mathcal{H}^{-1}(y) - \alpha(y)) =V_0 +  T_3\Phi_-.
\end{align}
Here $V_0$ is a constant extracted from $\Phi_-$, since $\Phi_-$ as defined in \eqref{fluxcom} is invariant under constant shifts, so that now $\Phi_-$ doesn't include a constant term. 
For an exact ISD solution, $\Phi_-=0$, and so the D3-brane feels no potential in this slow roll limit. The leading order potential then comes from the $\mathcal{O}(\delta)$ corrections to $\Phi_-$ coming from the gluing of the warped throat to the bulk Calabi-Yau.

The other regime, in which the brane is moving relativistically, is called DBI inflation. In this case, the motion is constrained by the causal speed limit, set by the positivity inside the square root in \eqref{lagrangianwarp}. This can lead to inflation, even if the potential is steep, because of the noncanonical kinetic terms. In this work we only consider the slowly rolling regime, and defer investigation of the DBI limit to a future investigation.

\subsubsection{Natural Inflation from a D3-brane?}
In this section we show our first attempt at realising Natural Inflation using a slow rolling D3-brane probe. We will find that we can't have a Planckian decay constant in a consistent manner. In Subsection~\ref{d5subsection} where we consider a probe wrapped D5-brane with flux, we'll find instead that we can consistently choose a Planckian decay constant.

Consider the WRC metric restricted to the $(\rho,\theta)$ subspace 
\begin{align}
ds^2 &= u^2\left[9\left(\frac{\rho^2 + 2/3}{\rho^2 + 1} \right)d\rho^2 + \left(\frac{3}{2}\rho^2 + 1\right)d\theta^2\right].
\end{align}
Now we make a coordinate transformation to canonical coordinates $(Z,\Theta)$, so that the kinetic terms appearing in the D3-brane slow roll Lagrangian \eqref{lagrangianwarp}  will be canonical 
\begin{align}
\frac{1}{2} T_3 \, \tilde{g}_{mn}\dot{y}^m\dot{y}^n 
&\approx \frac{1}{2}T_3u^2\left[9\left(\frac{\rho^2 + 2/3}{\rho^2 + 1} \right)\dot{\rho}^2 + \left(\frac{3}{2}\rho^2 + 1\right)\dot{\theta}^2\right] \label{DBIexp}
\\
&= \frac{1}{2}\dot{Z}^2+ \frac{1}{2}\dot{\Theta}^2.
\end{align} 
In the small $\rho$ limit, near the tip, the desired coordinates are given by 
\begin{align}
\rho &= \frac{1}{u\sqrt{6T_3}}Z
\\
\theta &= \frac{1}{u\sqrt{T_3}}\Theta.
\end{align}
We now examine the potential \eqref{totalpot} for the D3-brane, with our choice of homoegenous solutions to the Laplace equation, in terms of these canonical coordinates. It is
\begin{align}
\begin{split}
V(Z,\Theta) = V_0 + T_3 \bigg [ &\frac{ m^2 Z^2}{6 u^2T_3} + \frac{2 a_0u^2 T_3 }{3 Z^2} - a_0 \log \left(1+\frac{2  u^2T_3}{3 Z^2}\right) 
\\ 
&+\frac{3b_1}{4}  \left(4+\frac{9 Z^2}{ u^2 T_3}\right) \cos \left(\frac{\Theta }{ u\sqrt{T_3}}\right) \bigg ].
\end{split}
\end{align}
This will have a stable minimum in $Z$ at $Z_{min}$. If inflation begins at $Z$ close to $Z_{min}$ and sufficiently off to the side of the ridge along the $\Theta$ direction, the motion will be mostly in the $\Theta$ direction. We then identify $f = u\sqrt{T_3}$ for a D3-brane.

We now investigate whether one can arrange for the decay constant $ f = u\sqrt{T_3}$ to be of order $5M_p$.  Note that in our conventions, $T_3 = [(2\pi)^{3}g_sl_s^4]^{-1}$, and $\kappa_{10} = \frac{1}{2}(2\pi)^{4}g_sl_s^4$. We require 
\begin{align}
25M_p^2 &= u^2 T_3 =  \frac{u^2}{(2\pi)^{3}g_sl_s^{4}} , \label{plnkrange}
\end{align}
but we also have that the reduced Planck mass is related to the warped volume of the Calabi-Yau by $M_p^2 = V^w_6\kappa_{10}^{-2}$. We take the throat length to be $r_{\text{UV}}\gg r_{\text{min}}$. Throughout the region $r_{\text{min}}\ll r<r_{\text{UV}}$, the space is approximately $AdS_5 \times T^{1,1}$ and the warp factor goes like $\mathcal{H}\sim L_{T^{1,1}}^4/\rho^{4}$, where  $L_{T^{1,1}}^4=(27/4)\pi g_s N l_s^4$.  Under the assumption that $V_{throat}^w \gae V_{bulk}^w$, we get
\begin{align}
M_p^2 &\gae \ \kappa_{10}^{-2}\text{vol}({T^{1,1}})\int_{r_{\text{min}}}^{r_{\text{UV}}}y^5\mathcal{H}(y) dy
\approx  \frac{Nr_{\text{UV}}^2}{2(2\pi)^4g_sl_s^4}. \label{plnkvol}
\end{align}
Near the tip of the WRC the Planck mass receives a contribution
\begin{align}
\frac{Nr_{\text{min}}^2}{g_sl_s^4},
\end{align}
which will only contribute to \eqref{plnkvol} by multiplication of an order one factor  for $r_{\text{min}} \lae r_{\text{UV}}/50$. This is because in the small $\rho$ part of the WRC throat, where we are near to the $N$ D3-branes, the space is locally $AdS_5\times S^5$ and the warp factor is $\mathcal{H}\sim L_{S^5}^4/y^{4}$, now with $y$ the distance to the North Pole, where the $N$ D3's are located. 

In order to match \eqref{plnkrange} to \eqref{plnkvol} we must take the throat length $r_{\text{UV}}$ to be
\begin{align}
r_{\text{UV}}^{2} \approx \frac{4\pi u^2}{25N}
\ll u^2, \text{ for large } N \gg 1.
\end{align}
But $u$ is the natural length scale of the RC. To have the length of the throat $r_{\text{UV}}$ hierarchically smaller than $u$ seems unnatural, since we assumed a very long throat for the noncompact limit, and also for the $M_p$ approximation coming mainly from the throat. It seems we can't consistently choose $f=5M_p$ for a D3-brane.

Finally, for a later comparison with the D5-brane case, we note that turning on a constant electric flux of $\epsilon<1$ on the D3-brane gives a factor $(1-\epsilon^2)^{1/2}$ in front of the DBI part of the action, with the CS part of the D3 action left unchanged. This leads to a non-cancellation of the inverse warp factor, and so the following potential term appears
\begin{align}
V = T_3\left [(1-\epsilon^2)^{1/2} - 1\right ] \mathcal{H}^{-1}.
\end{align}
But we note that this comes with a negative sign, and so acts to destabilize the overall potential,  making it unsuitable for achieving a stable minimum in $Z$. In the D5-brane case, this non-cancellation will come with the opposite sign, as then the flux appears in the CS action.

\subsection{Wrapped D5-brane in the WRC} \label{d5subsection}
In this section we'll find that for the probe wrapped D5-brane with flux, we can obtain $f\approx5M_p$ within the long throat approximation. In this case the non-cancellation of the DBI and CS terms works to our advantage, giving a quadratic term in the potential with a positive coefficient.

We consider the same WRC background but this time we place a probe D5-brane in it, with 4 of it's dimensions extended along the 4 noncompact spacetime directions and wrap the remaining two spatial dimensions around a 2-cycle $\Sigma_2$ inside the compact space $p$ times. We also turn on an $F_2$ flux on the D5-brane through $\Sigma_2$.

The action for the $p$-wrapped D5-brane with worldvolume coordinates $\xi^{\alpha}$ and worldvolume $W_5$ is 
\begin{align}
S_{D5} &= S_{DBI-D5} + S_{CS-D5} 
\\
\begin{split}
= &-T_5\int_{W_5}d^6\xi\sqrt{-\det(P_6[g_{MN} + B_{MN} + 2\pi\alpha'{F}_{MN}])}
\\
& + T_5\int_{W_5} P_6\left [ C_6 + C_4 \wedge (B_2 + 2\pi\alpha'F_2) \right  ]
\end{split}
\label{wd5}
\end{align}
where $P_6$ is the pullback of a 10D tensor to the 6D brane worldvolume, and $T_5 = [(2\pi)^5g_sl_s^6]^{-1}$. 

It's important to distinguish between the \textit{embedding} of the D5-brane in the 10D spacetime, and the \textit{wrapping} of the D5 on a 2-cycle $\Sigma_2$ in the 6D internal space.

The embedding is a relation between the brane worldvolume coordinates $\xi^{\alpha}$ and the 10D spacetime coordinates $x^M$, given as $\xi^{\alpha} = k^{\alpha}(x^M), \alpha=0,...,5$ for some function $k^{\alpha}$. Similarly to \cite{Becker2007}, we choose the simple embedding
\begin{align}
\xi^{\alpha}   = (x^0,x^1,x^2,x^3,\theta_1,\phi_1).
\end{align}

We specify the wrapping on the 2-cycle $\Sigma_2$ in the 6D internal space by taking $6-2=4$ relations on the internal $y^m$ coordinates 
\begin{align}
r=\text{constant}
&&
\theta_2= f(\theta_1) = -\theta_1
\\
\psi =\text{constant}
&&
\phi_2= g(\phi_1) = -\phi_1.
\end{align}
With this choice of embedding and wrapping, the pullback of the 10D metric $g_{MN}$ to the 6D D5-brane worldvolume is
\begin{align}
P_6[g]_{\alpha\beta} &= \frac{\p X^M}{\p \xi^{\alpha}}\frac{\p X^N}{\p \xi^{\beta}}g_{MN},
\end{align}
which takes the following diagonal form
\begin{align}
P_6[g]_{00} &= -\mathcal{H}^{-1/2}(1-\mathcal{H}v^2)
\\
P_6[g]_{ii} &= a^2\mathcal{H}^{-1/2}
\\
P_6[g]_{\theta_1\theta_1} &= \frac{1}{3}\mathcal{H}^{1/2}(r^2 + 3u^2)
\\
P_6[g]_{\phi_1\phi_1} &= \frac{1}{3}\sin^2 \theta_1\mathcal{H}^{1/2}(r^2 + 3u^2),
\end{align}
where there's no summation implied on the $ii$ component, and we restrict to only have motion in the $r,\theta_2$ directions, so that the speed squared of the brane is
\begin{align}
v^2 = \left (\frac{r^2 + 6u^2}{r^2 + 9u^2}\right )\dot{r}^2 + \frac{1}{6}\left (r^2 + 6u^2\right )\dot{\theta}_2^2.
\end{align}
We choose to turn on a worldvolume flux $F_2$, of strength $q$ along the wrapped 2-cycle, so that its pullback has the following non-zero components
\begin{align}
P_6[2\pi\alpha'F_2]_{\theta_1\phi_1} = 2\pi\alpha'\frac{q}{2}\sin\theta_1 = -P_6[2\pi\alpha'F_2]_{\phi_1\theta_1}.
\end{align}

As an aside, we note that we have chosen to turn on an $F_2$ worldvolume flux, but we could also have turned on a $B_2$ worldvolume flux, as discussed in Appendix~\ref{appendix:d5b2}. The wrapped D5-brane would lead to a potential for the $b$-axion associated with integrating this $B_2$ over the wrapped 2-cycle. This $b$-axion was used to inflate from in models of axion monodromy inflation \cite{Silverstein:2008sg}. However, for suitable initial conditions on the size of $b$, this term will not affect our inflationary dynamics from the position modulus of the wrapped D5-brane. 

Now we have the following term in the DBI part of the action
\begin{align}
S_{DBI-D5} &= -pT_5\int_{{M}_4 \times \Sigma_2} d^4xd\theta_1 d\phi_1\sqrt{-\det(P_6[g + 2\pi\alpha'F_2])} 
\\
&= -pT_5\int_{{M}_4} d^4xa^34\pi\mathcal{H}^{-1}\mathcal{F}(r,\theta_2)^{1/2}\sqrt{1-\mathcal{H}v^2} \label{midd5act}
\\
\text{where } \mathcal{F}(r,\theta_2) &\equiv \frac{\mathcal{H}}{9}(r^2 + 3u^2)^2 + (\pi\alpha' q)^2.
\label{sdbi}
\end{align}
Now we need to do a slow roll expansion, and check what terms multiply the kinetic terms, and then define new coordinates which have canonical kinetic terms. In the 4D Lagrangian density, (with $a^3$ absorbed into the integration measure) the coefficient of $\dfrac{1}{2}v^2$ in the expansion of the square root is
\begin{align}
4\pi pT_5 \mathcal{F}(r,\theta_2)^{1/2} . 
\label{sqrt}
\end{align}

 Since the open string modulus has mass dimensions, we expect this modulus to be affected by the warping. In the case of the D3-brane, the warp factors exactly cancelled in \eqref{lagrangianwarp} to give \eqref{lagrangianwarpexpand} with 4D canonical fields with no powers of the warp factor. However, in the case of a wrapped D5-brane, we see this cancellation no longer occurs, with dependence on the warp factor to the power of $1/2$ as set by $\mathcal{F}^{1/2}$ in \eqref{midd5act}. 

If we neglect $\mathcal{O}(\alpha'^2 q^2)$ and take the AdS limit of the throat,  we would get $\mathcal{F}\sim R^4/9$ giving the same result as found in \cite{Becker2007}.
However, we are interested in the \textit{small} $r$ region of the throat, near the stack, where we will fix $r=r_{\text{min}} \ll r_{\text{UV}}$. 

In contrast to the D3-brane case, we can now take $r_{\text{UV}} \sim u$, so that the length of the throat is of order the resolution parameter, which is the natural lengthscale of the WRC geometry. This is because we have more freedom in the model from the wrapping number $p$. 

For the moment let's take the following assumption on the size of the flux $q$
\begin{align}
q \lae \frac{1}{\pi}\left (\frac{u}{r_{\text{min}}}\right )^2\sqrt{4\pi g_s N}, \label{qfapprox}
\end{align}
which means that to leading order in $u/r_{\text{min}}$, $\mathcal{F}$ has behaviour   
\begin{align}
\mathcal{F}(r,\theta_2)^{1/2} \approx \frac{u^2L_{S^5}^2}{r_{\text{min}}^2}  \approx  \left (\frac{u}{r_{\text{min}}}\right )^2 l_s^2\sqrt{4\pi g_s N} \label{Fapprox}
\end{align}
with the $q^2$ term possibly contributing only an $\mathcal{O}(1)$ numerical factor in front of this. Here $L_{S^5}^4=4\pi g_s Nl_s^4$ is the fourth power of the $AdS_5\times S^5$ radius, which is the near stack geometry created by the $N$ D3's. 

\subsubsection{Decay Constant $f$}
The canonical kinetic coordinate $\Theta$ is
\begin{align}
 {\Theta} \equiv  \frac{u^2}{r_{\text{min}}}  (4\pi pT_5L_{S^5}^2)^{1/2} {\theta_2}. \label{Theta}
\end{align}

The dominant contribution to the Planck mass from a long warped throat of length $r_{\text{UV}}\sim u$ is  
\begin{align}
M_p^2 &\gae \ \kappa_{10}^{-2}\ \text{vol}({T^{1,1}})\int_{0}^{u}y^5\mathcal{H}(y)
\approx \frac{Nu^2}{2(2\pi)^4g_sl_s^4}. \label{plnkvolu}
\end{align}

We want the decay constant $f$ to be $5M_p$ for observable tensor modes, which occurs for the canonical field $\Theta$ in \eqref{Theta} when we set the coefficient of $\theta_2^2$ to be equal to $25 M_p^2$. Using the reduced Planck mass from the volume of the throat \eqref{plnkvolu}, we find that this requires
\begin{align}
p \sim \frac{25}{4\pi}\left (\frac{r_{\text{min}}}{u}\right )^2\sqrt{\frac{N\pi^3}{4\pi^2 g_s}}. \label{pd5}
\end{align}
Note that there is dependence on the ratio $(r_{\text{min}}/u)^2$, which is small for $r_{\text{min}}$ near the tip. This will be helpful for keeping the backreaction under control, as we will show in Appendix~\ref{backreactionappend}. 

\subsubsection{Scale of Inflation}
We now wish to set the scale of inflation to be $M_{\text{GUT}}$.
Writing the action for the D5-brane minimally coupled to gravity in an FRW spacetime and expanding in slow roll fields gives
\begin{align}
S_{D5} &= 
\int_{{M}_4} d^4x\sqrt{g_{\text{FRW}}}\left [ \frac{M_p^2}{2}\mathcal{R}_4 + \frac{1}{2}\dot{Z}^2 + \frac{1}{2}\dot{\Theta}^2  - V \right ]
\\
\text{where } V &= \frac{M_p^2}{4}\left [ \varphi(y) + \lambda\Phi_- \right ]  \label{splitaction} 
\end{align}
and
\begin{align}
\Phi_- &= \overline{\Phi}_- + \Phi_h
\\
\varphi(y) &= \frac{4}{M_p^2}4\pi pT_5 \mathcal{H}^{-1}\left ( \mathcal{F}^{1/2} - l_s^2\pi q  \right )
\qquad
\\
\lambda &=  \frac{4}{M_p^2}4 \pi^2 l_s^2 T_5 p q . \label{lambda}
\end{align}

We are aiming for a Natural Inflation type potential, which we derive from the homogeneous solution \eqref{homd5}. We want the cosine term to be
\begin{align}
V = M_{GUT}^4 \cos \left (\frac{\Theta }{5M_p}\right )\label{natpothom}.
\end{align}
We now ask what value of $q$ is required to match the coefficient of $\cos (\Theta/5M_p)$ in \eqref{splitaction}  to be $M_{GUT}^4$, in the small $\rho$ limit. From \eqref{splitaction}, \eqref{lambda} and \eqref{homd5}, this requires
\begin{align}
4 \pi^2 l_s^2 T_5 p q 3b_1 &= M_{GUT}^4.
\end{align}
Taking $b_1 = \delta B$, with $B=\mathcal{O}(1)$, the LHS becomes
\begin{align}
4 \pi^2 l_s^2 T_5 pq  3b_1 
&\approx 2.1 \times 10^{-2}\frac{N^{1/2}}{g_s^{3/2} l_s^4}\left ( \dfrac{r_{\text{min}}}{u}\right )^2 \delta B q.
\end{align}
The GUT scale is 
$M_{\text{GUT}} =  10^{16} \text{GeV} \approx 4\times 10^{-3}M_p$
and using the reduced Planck mass from \eqref{plnkvolu} gives
\begin{align}
M_{\text{GUT}}^4 
\approx 2.6 \times 10^{-17}\frac{N^2u^4}{g_s^2l_s^8},
\end{align}
meaning we need to take $q$ of order
\begin{align}
q \approx 1.2 \times 10^{-15} \frac{N^{3/2}}{g_s^{1/2} \delta B} \left (\frac{u}{l_s}\right )^4 \left (\frac{u}{r_{\text{min}}}\right )^2 \label{qgut}
\end{align}
to get the desired GUT scale.

\subsubsection{Backreaction of a Wrapped D5-brane} \label{backreactionsec}
Unlike the D3-brane, the D5-brane can backreact on both the warp factor and the internal geometry. The backreaction on the internal geometry is possibly lethal to the assumption that the leading order terms in the potential are small, coming from the gluing of the warped throat to the CY. We should check that our chosen values for $p$, in \eqref{pd5}, and $q$, in \eqref{qgut}, are small enough so that backreaction effects are negligable. 

Table~\ref{tab:parameters} summarizes the data we take for various parameters. The model is fairly robust to small changes in these values.

\begin{table}[h]
\centering
\begin{tabular}{| l | c | c | c | c | c | c | r |}
  \hline  
  Parameter & $B$ & $g_s$ & $N$ & $u$ & $\delta$ & $r_{\text{min}}$ & $\sin \theta_2$ \\
  \hline  
Data & $1$ & $10^{-1}$ & $10^4$ & $50 l_s$ & $10^{-2}$ & $l_s$ & $10^{-2}$ \\
  \hline  
\end{tabular}
 \caption{Compactification data for our model}
  \label{tab:parameters}
\end{table}

Before we check that the backreaction from the wrapped D5-brane is under control, we can first check the value of $q$, given in \eqref{qgut} for the scale of inflation to be the GUT scale, against the constraint we already imposed on it in \eqref{qfapprox} for the approximation of $\mathcal{F}^{1/2}$ in \eqref{Fapprox}.
We require
\begin{align}
1.2 \times 10^{-15} \frac{N^{3/2}}{g_s^{1/2} \delta B} \left (\frac{u}{l_s}\right )^4 \left (\frac{u}{r_{\text{min}}}\right )^2 &\lae \frac{1}{\pi}\left (\frac{u}{r_{\text{min}}}\right )^2\sqrt{4\pi g_s N}
\\
\Leftrightarrow \frac{N}{B g_s \delta}\left (\frac{u}{l_s}\right )^4 &\lae  9.4\times 10^{14}
\end{align}
independent of the value of ${r_{\text{min}}}$. For the data in Table~\ref{tab:parameters}, we have
\begin{align}
\frac{N}{B g_s \delta}\left (\frac{u}{l_s}\right )^4 &\sim 6.25 \times 10 ^{13} \lae   9.4\times 10^{14} \label{fapproxcorrect}
\end{align}
as required.

In Appendix~\ref{backreactionappend} we show further that for this choice of data, the backreaction is under control. We calculate the size of the backreaction on the warp factor and on the internal geometry, and find that they are small compared to what is produced by the stack. 

\subsubsection{The Stable Radial Minimum}
We now confirm that we can find a stable radial minimum in $\rho$ for our potential.
Using our chosen $p$ in \eqref{pd5} and $q$ in \eqref{qgut}, and the data in Table~\ref{tab:parameters} we get
\begin{align}
\varphi(y) &= \frac{4}{M_p^2}4\pi u^2pT_5\mathcal{H}^{-1/2}  \approx \frac{180}{u^2}\rho^2
\\
\lambda &=  \frac{4}{M_p^2}4 \pi^2 l_s^2 T_5 p q \approx 4 \frac{M_{\text{GUT}}^4}{\delta M_p^2} \approx \frac{0.2}{\delta u^2} \approx \frac{20}{ u^2}
\end{align}
where we've used that $M_{\text{GUT}} = 4\times 10^{-3}M_p$, and 
\begin{align}
M_p^2 &= \frac{Nu^2}{2(2\pi)^4g_s l_s^4} \approx\frac{ 2 \times 10^{8}}{u^2}
\end{align}
for our data.
Then we have
\begin{align}
\lambda M_p^2/4 \approx 100 M_{\text{GUT}}^4. \label{lambdamp}
\end{align}
We note that in the potential there is the term
\begin{align}
\frac{M_p^2}{4}\varphi(\rho) \approx M_{\text{GUT}}^4 880 \rho^2 \label{vpot}
\end{align}
coming from the non-cancellation of DBI and CS terms.
The leading order behaviour of the potential for small $\rho$ is then
\begin{align}
V &\approx  M_{\text{GUT}}^4\left [ 880\rho^2   + \frac{{A_0}}{\rho ^2}+{2A_0} \log \rho +3B_1 \cos \left (\frac{\Theta }{5M_p}\right )\right ].
\end{align}
where we have kept the $\rho^{2}$ term from $\varphi(y)$, since it comes with a large coefficient, but neglected positive powers of $\rho$ from $\Phi_h$ in the small $\rho$ limit.

We now look for a minimum $\rho_{\text{min}}$ small enough that our selected data $r_{\text{min}} \sim u/50$ is valid. This means that we require $\rho_{\text{min}} \sim 1/150 \approx 7 \times 10 ^{-3}$.

This requires us taking $A_0 \approx 10^{-5}$, which gives us $\rho_{min} \approx  10 ^{-2}$. This is a bit of fine-tuning, but the only restriction we have on the $a_i$ is that they are less than $ \mathcal{O}(\delta) = 10^{-2}$. The fact that we've taken all the other $a_i$ as zero is at least consistent with having a very small $a_0$. 

\begin{figure}[tbp]
    \centering
    \includegraphics[width=0.8\textwidth]{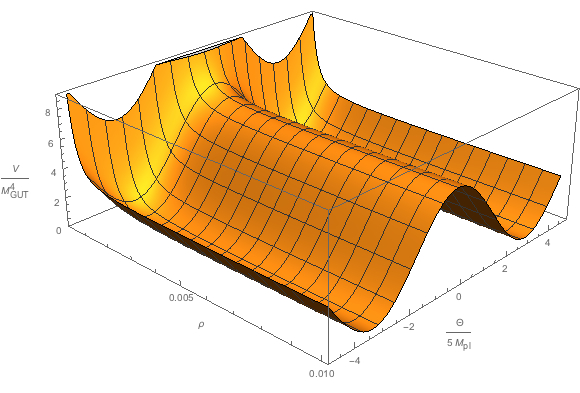} 
    \caption{Natural Inflation potential for the D5-brane}
    \label{fig:pot}
\end{figure}

Once we have stabilized in the $\rho$ direction at $\rho_{min}$, the effective potential is only in the $\Theta$ direction, and has the form of the Natural Inflation potential.  A plot of the potential, with this  value of $A_0$ and $B = 1$  is shown in Figure~\ref{fig:pot}.

We could include the correction from the Ricci scalar, calculated in Appendix~\ref{ricciappend}, which contributes at leading order with a quartic term in $\rho$ with a large coefficient. This changes the potential to 
\begin{align}
V &\approx  M_{\text{GUT}}^4\left [ 880\rho^2 + 4500\rho^4   + \frac{{A_0}}{\rho ^2}+{2A_0} \log \rho +3B_1 \cos \left (\frac{\Theta }{5M_p}\right )\right ]. \label{potcorrected}
\end{align}
However, the position of the minimum is highly insensitive to the addition of this term in the small $\rho$ limit. A priori, one might have expected corrections to contribute towards the position of the minimum, but it seems the dominant feature is the small value of $A_0$. 

\section{Conclusion} \label{sec:conc}

In this paper we have investigated D-brane potentials in the background of the warped resolved conifold (WRC) and applied this to give a model of Natural Inflation.  
The potentials arise as perturbations to the ISD solution from the gluing of the warped throat to a bulk Calabi-Yau.
These perturbations are the solutions to the Laplace equation on the unwarped resolved conifold (RC). We know the exact solutions to this equation, valid anywhere within the throat, in particular at the tip. This is not the case for the warped deformed conifold geometry, in which the solutions to the Laplace equation on the deformed conifold are only valid in the mid-throat region, far from the tip.

This allowed us to exploit the effect of the warping, which is strongest at the tip. 
We inflated using a periodic angular coordinate which had a potential involving a cosine of this coordinate, giving us a model of Natural Inflation. We now summarise how we achieved a Planckian decay constant for our Natural Inflation potential, given the difficulty this has posed in previous attempts at embedding natural inflation in string theory. 

A crucial ingredient in this respect is the choice of inflating with a D5-brane rather than a D3-brane. Using first a D3-brane, we found that we couldn't obtain a Planckian decay constant. Increasing the number of D3's increases the decay constant but the amount required has been shown to yield a large backreaction \cite{Pajer:2008uy}. 

We then considered instead a wrapped D5-brane probe, with electric flux turned on along the wrapped directions. We found it was possible to get a Planckian decay constant for this probe and simultaneously set the energy scale of inflation to be the GUT scale, whilst maintaining control over the backreaction and supergravity approximation. 
This is because the pullback of the DBI action to the D5-brane worldvolume produces a dependence on the warp factor through the term $\mathcal{F}^{1/2}$, which is not present in the case of the D3-brane.
 
The term $\mathcal{F}$ is proportional to the warp factor and by setting the initial conditions such that the D5-brane begins at a small radial displacement compared to the lengthscale $u$, allows for $\mathcal{F}^{1/2}$ to be large, as emphasised in \eqref{Fapprox}. Since the decay constant $f$ is proportional to $p \mathcal{F}^{1/2}$, as shown in \eqref{sqrt}, one can achieve a Planckian value for the decay constant by either a large $p$ or a large $\mathcal{F}^{1/2}$. In our case we choose $\mathcal{F}^{1/2}$ to be large, as it has already been shown that choosing a large $p$ would lead to a large backreaction as the brane would become very heavy. Having some moderate wrapping $p>1$ is helpful however, since in the case of no wrapping a Planckian decay constant can only be achieved for an extreme hierarchy of $r_{min} \ll u$, which would be difficult to achieve. The choice of moderate $p$ and the initial radial position of the D5-brane near the tip allows for a Planckian $f$.


The presence of 2-form flux $q$ is not crucial for the Planckian $f$, as it does not enter directly into the expression for $f$, as long as we use the
$AdS_5\times S^5 $  approximation valid near the stack of branes at the tip. However, $q$ does appear in the CS action for the D5-brane, and so a non-zero $q$ is crucial in order to get a cosine potential term in the first place, coming from the solution to the Laplace equation on the resolved conifold. By \eqref{lambda}, this potential is proportional to $q$ and so this can be chosen to set the overall energy scale of the natural inflation potential to be $M_{GUT}$.

We found that we could tune one coefficient of a solution to the Laplace equation to arrange for a stable minimum in the radial direction which is very close to the tip. The slope of the potential was found to be much greater in the radial direction than in the angular direction. We can arrange for inflation to begin once the brane has stabilized in the radial direction, with inflation solely along the angular direction. Enough e-folds are produced if inflation begins near the top of the potential in the angular direction.

Estimates of the backreaction for the case of a wrapped D5-brane were presented in Appendix~\ref{backreactionappend} where it was shown that there is a choice of stringy parameters for which the supergravity and probe brane approximations are valid. 
In Appendix~\ref{ricciappend}, we investigated corrections to the potential when the noncompact limit approximation is relaxed, to include the effects of a nonzero 4D Ricci scalar. This involved solving the Poisson equation on the RC, using the Green's function method outlined in
 \cite{2006JHEP...11..031B}. The leading order term only contributed at fourth order in the small radial coordinate, subleading to other terms in the potential. It had a very small effect on the value of the radial coordinate at the minimum. 

Future work could explore more general motion in both the radial and the other angular directions in the WRC.  We focussed in this paper on slow roll inflation in an angular direction near the tip of the WRC.
This could also be extended to the case of relativistic branes and to a DBI spinflation scenario on the WRC. 
Given that one knows the explicit form of the solutions of the Laplace equation and the Green's function on the whole of the RC, one could easily broaden this study to explore other regions of the RC throat.

We have not presented how to stabilize the closed string moduli within the RC throat, as it can't support non-trivial (2,1)-form flux, required to stabilize the bulk complex structure moduli and the axiodilaton, whilst preserving $\mathcal{N}=1$ supersymmetry. Future work could investigate complex structure moduli stabilization in the RC - in particular, supersymmetry breaking (3,0)-fluxes may present a way forward, and may lead to supersymmetry broken at a high scale. One could also ask how K\"{a}hler moduli stabilization is affected, and its cosmological implications. This could also impact on the stabilization of the axiodilaton as well. In addition, the idea of the decoupling of closed string moduli from open string moduli could be pursued. 

Finally, an interesting broad question arising from this work is whether observable values of $r$ can be achieved in general from D3-brane inflation and to what extent one is forced to consider objects such as wrapped D5-branes or other brane constructions in order to produce a large value of $r$ within a warped throat setup.

\acknowledgments
We would like to thank John Ward and David Mulryne for useful discussions and comments. The work of ZK was supported by an STFC studentship. The research of ST was supported by STFC consolidated grants ST/J000469/1 and ST/L000415/1.

\appendix
\numberwithin{equation}{section}
\section{Backreaction for the D5-brane} \label{backreactionappend}
The D5-brane can backreact on both the warp factor and the internal geometry, so we need to estimate the size of each. Our strategy will be to first assume the backreaction on the internal geometry is small, so that we have an ISD solution with warp factor $\mathcal{H}$. We then compute the backreaction on this warp factor. In finding that this is small, we use the non-backreaction of the warp factor to compute the possible backreaction on the internal geometry. We will find that this is small too, making our approach self-consistent.

We should begin with the full action for the SUGRA background and include terms in the action for all localized sources, which include the stack of $N$ D3-branes located at the north pole of the finite $S^2$, with no fluxes turned on, as well as the mobile probe wrapped D5-brane with flux. The full action is then
\begin{align}
\begin{split}
S &= \frac{1}{2\kappa_{10}^{2}}\tilde{S}
- NT_3\int_{{M}_4}d^4\chi\sqrt{-\det(P_4[g_{MN} ])} + NT_3\int_{{M}_4}P_4[C_4]
\\
&- T_5\int_{{M}_4 \times \Sigma_2}d^6\xi\sqrt{-\det(P_6[g_{MN} + \mathcal{F}_{MN}])} + T_5\int_{{M}_4 \times \Sigma_2}P_6\left [ C_6 + C_4 \wedge \mathcal{F}_2 \right  ]
\end{split}
\end{align}
where $\tilde{S}$ is given by \eqref{IIBaction}.
 In our WRC geometry, $C_6 =0$, $B_2 = 0$ and we've turned on the flux $F_2 = \frac{q}{2}\sin\theta_1 d\theta_1 \wedge d\phi_1$ on the D5-brane. 

We now need to promote the worldvolume integrals to integrals over the full $M_{10}$ space, in order to vary this action and get the equations of motion. To do this we introduce the following D3 charge densities
\begin{align}
{\rho}_3^{ND3} &= \frac{N}{\sqrt{{g}_6}}\delta(r)\delta(\psi)\delta(\theta_2)\delta(\phi_2)
\delta(\theta_1)\delta(\phi_1)
\\
{\rho}_3^{pD5} &= \frac{p}{\sqrt{{h}_4}}\delta(r-r^*)\delta(\psi-\psi^*)\delta(\theta_2-\theta_2^*)\delta(\phi_2-\phi_2^*)
\\
{\rho}_3^{pqD5} &= \frac{p}{\sqrt{{g}_6}}(2\pi\alpha')\left (\frac{q}{2}\sin \theta_1\right )\delta(r-r^*)\delta(\psi-\psi^*)\delta(\theta_2-\theta_2^*)\delta(\phi_2-\phi_2^*)
\end{align}
so that the stack of $N$ D3's are at the tip, $y=0$, and the p-wrapped probe D5 is at $(r^*,\psi^*,\theta_2^*,\phi_2^*)$. The metric $h_4$ is the warped metric on the 4D space transverse to the brane in the extra 6 dimensions. We then define the 6D unwarped densities (with a tilde) via
\begin{align}
\tilde{\rho}_3^{ND3} = \mathcal{H}^{3/2}{\rho}_3^{ND3} 
\qquad
\tilde{\rho}_3^{pqD5} = \mathcal{H}^{3/2}{\rho}_3^{pqD5}. \label{pqdensity}
\qquad
\text{ since }\sqrt{g_{6}}=\mathcal{H}^{3/2}\sqrt{\tilde{g}_6}.
\end{align}
For the 4D density, we have
\begin{align}
\tilde{\rho}_3^{pD5} = \mathcal{H}{\rho}_3^{pD5} \text{ since }\sqrt{h_{4}}=\mathcal{H}\sqrt{\tilde{h}_4}.
\end{align}
Note that the Hodge dual $\star_6$ of the warped 6D metric $g_6$ acts on a 6D warped density ${\rho}$ as 
\begin{align}
\star_6 {\rho} &={\rho} \sqrt{g_6} \  dr \wedge d\psi \wedge d\theta_1 \wedge d\phi_1 \wedge d\theta_2 \wedge d\phi_2 .
\end{align}
We then have the full 10D action
\begin{align}
\begin{split}
S &= \frac{1}{2\kappa_{10}^{2}}\tilde{S}
- T_3\int_{M_{10}} d^{10}x\sqrt{-\det(P_4[g_{MN} ])}\sqrt{{g}_6}\rho_3^{ND3} + T_3\int_{M_{10}}C_4 \wedge \star_6 \rho_3^{ND3} 
\\
&- T_5\int_{M_{10}} d^{10}x\sqrt{-\det(P_6[g_{MN} + \mathcal{F}_{MN}])}\sqrt{{h}_4}\rho_3^{pD5} + T_5\int_{M_{10}}C_4 \wedge \star_6 \rho_3^{pqD5}.
\end{split}
\end{align}

We obtain the stress tensors from the DBI part of each local brane action. For the stack of $N$ D3-branes we get the following non-zero components for the stress-energy tensor
\begin{align}
T^0_0 = T^1_1 = T^2_2 = T^3_3 &= - T_3\rho_{3}^{N D3}, \label{d3stress}
\end{align}
so that 
\begin{align}
\left (T^m_m - T^{\mu}_{\mu} \right )_{ND3} &= 4 T_3\rho_{3}^{N D3}
\label{d3trace}
\end{align}
where we've used the shorthand
\begin{align}
T^m_m - T^{\mu}_{\mu}  \equiv \sum_{M=4}^{9} {{T}_M}^M - \sum_{M=0}^{3} {{T}_M}^M.
\end{align}
Now we do a similar calculation for the p-wrapped D5-brane. Neglecting the $\mathcal{O}(\alpha'^2q^2)$ flux contribution, we get the following non-zero components for the stress-energy tensor
\begin{align}
T^0_0 = T^1_1 = T^2_2 = T^3_3 = T^{\theta_1}_{\theta_1} =  T^{\phi_1}_{\phi_1}  &= - T_5\rho_{3}^{pD5}
\\
\text{so that }
\left (T^m_m - T^{\mu}_{\mu}\right )_{pD5} &= 2 T_5\rho_{3}^{pD5}.
 \label{d5trace}
\end{align}

Varying the full 10D action with respect to $C_4$ gives the Bianchi identity 
\begin{align}
d \tilde{F}_5 = H_3 \wedge F_3 + 2 \kappa_{10}^2\left ( T_3 \star_6 \rho_3^{ND3} + T_5\star_6 \rho_3^{pqD5} \right ).
\end{align}
Using the warped background ansatz, this becomes
\begin{align}
\tilde{\nabla}^2  \alpha &= ie^{2A}\dfrac{G_{mnp}\star_6\overline{G}^{mnp}}{12\text{Im}\tau} + 2e^{-6A}\p_m \alpha \p^m e^{4A} + 2 \kappa_{10}^2e^{2A}\left ( T_3  \rho_3^{ND3} + T_5 \rho_3^{pqD5} \right ). \label{bianchialpha}
\end{align}
The trace of the Einstein equations can  be written \cite{Giddings2002}
\begin{align}
\begin{split}
\tilde{\nabla}^2  e^{4A} &=  \frac{\kappa_{10}^2}{2}e^{2A}\left[ \frac{1}{4}(T^m_m - T^{\mu}_{\mu})_{ND3} + \frac{1}{4}(T^m_m - T^{\mu}_{\mu})_{pD5}   \right ]
\\
&+ e^{2A}\dfrac{G_{mnp}\overline{G}^{mnp}}{12\text{Im}\tau} + e^{-6A}(\p_m\alpha\p^m\alpha + \p_m e^{4A}\p^m e^{4A}) . \label{traceee}
\end{split}
\end{align}
Combining \eqref{bianchialpha} and \eqref{traceee} gives
\begin{align}
\begin{split}
\tilde{\nabla}^2  (e^{4A}-\alpha) &= \frac{e^{2A}}{24\text{Im}\tau}|iG_3 - \star_6G_3|^2      + e^{-6A}|\p(e^{4A}-\alpha)|^2 
\\
&+ 2 \kappa_{10}^2e^{2A}\left (\frac{1}{4}(T^m_m - T^{\mu}_{\mu})_{ND3} - T_3  \rho_3^{ND3} \right )
\\
&+  2 \kappa_{10}^2e^{2A}\left ( \frac{1}{4}(T^m_m - T^{\mu}_{\mu})_{pD5} - T_5 \rho_3^{pqD5} \right ). \label{bianchid5}
\end{split}
\end{align}
 The first two terms on the RHS are non-negative. The third term actually vanishes by \eqref{d3stress}, but note that this is a special result for D3-branes. 
The fourth term for the D5-brane doesn't vanish, so we need to work out its size. It can be written in terms of the warp factor as 
\begin{align}
2 \kappa_{10}^2 \mathcal{H}^{-1/2}\left ( \frac{1}{4}(T^m_m - T^{\mu}_{\mu})_{pD5} - T_5 \rho_3^{pqD5} \right ). \label{d5internal}
\end{align}
In order to estimate its size we need the warp factor. For the moment let's assume \eqref{d5internal} is small so that we can ignore it. We'll come back to the size of this term after we have computed the backreaction on the warp factor. This allows us to begin with the usual ISD solution
$
G_-=0 = \Phi_-
$.

\

{ \centerline{\textit{\textbf{Backreaction on the Warp Factor}}}}

\

We now compute the backreaction on the warp factor. We write the trace of Einstein's equations \eqref{traceee} in the form
\begin{align}
-\tilde{\nabla}^2  e^{-4A} 
&= 2{\kappa_{10}^2} T_3  \tilde{\rho}_3^{ND3} +  {\kappa_{10}^2}T_5\mathcal{H}^{1/2}\tilde{\rho}_3^{pD5}.    \label{sourcewarp}
\end{align}
We begin with the warp factor arising as the Green's function on the WRC for a stack of $N$ D3's placed at the north pole $\theta_2 =0$ of the $S^2$ at $r=0$ to get $\mathcal{H}$. The result near the tip $r=r_{\text{min}}$, but away from the north pole, is given by
\begin{align}
\mathcal{H} &= \frac{4\pi g_s N l_s^4}{r_{\text{min}}^4}.
\end{align}
We now consider corrections to this from modifying the Green's function equation to 
\eqref{sourcewarp}. 
We will compute the size of each term on the RHS of \eqref{sourcewarp}, and find that the dominant one is from the stack. 

We want to compare factors in front of dimensionless delta functions, and since $r$ has dimensions of length, we should look at dimensionless $\rho$, and use the scaling property of delta functions
$
\delta(r)  = \frac{1}{3u}\delta(\rho).
$
Let's define the following combinations of dimensionless delta functions
\begin{align}
\delta(M_6) &\equiv  \delta(\rho)\delta(\psi)\delta(\theta_2)\delta(\phi_2)
\delta(\theta_1)\delta(\phi_1)
\\
\delta(M_4^*) &\equiv  \delta(\rho-\rho^*)\delta(\psi-\psi^*)\delta(\theta_2-\theta_2^*)\delta(\phi_2-\phi_2^*).
\end{align}
Then the first term on the RHS of \eqref{sourcewarp} coming from the stack is
\begin{align}
2{\kappa_{10}^2} T_3  \tilde{\rho}_3^{ND3} = 
3(2\pi)^5\dfrac{Ng_sl_s^4}{r_{\text{min}}^3u^3\sin\theta_1\sin\theta_2}\delta(M_6).
\end{align}
The second term in \eqref{sourcewarp} coming from the $p$-wrapped D5-brane probe, evaluated at $r\approx r_{\text{min}} \ll u$ is
\begin{align}
{\kappa_{10}^2}T_5\mathcal{H}^{1/2}\tilde{\rho}_3^{pD5}  &= \frac{(2\pi)^3}{16}\frac{25Ng_sl_s^4}{r_{\text{min}}u^5\sin\theta_2}\delta(M_4^*), \label{psize}
\end{align}
using $T_5 = [(2\pi)^5g_sl_s^6]^{-1}$.
We see that the probe D5-brane source term for the warp factor is much smaller than that sourced by the stack of $N$ D3-branes, as long as
\begin{align}
  \left (\frac{r_{\text{min}}}{u} \right )^2 &\ll \frac{48}{25}(2\pi)^2\frac{1}{\sin\theta_1} \approx \frac{76}{\sin\theta_1},\label{backreactionwarp}
\end{align}
and since $\sin\theta_1 \leq 1$, the RHS of \eqref{backreactionwarp} is greater than $76$. So our $p$-wrapped D5-brane can be neglected in the warp factor equation since we have $r_{\text{min}}\ll {u} $. 

\

\centerline{ \textbf{\textit{Backreaction on the internal geometry}}}

\

Now that we've shown that the probe approximation for the warp factor is self-consistent, we can go back to the equation for the internal geometry to check the validity of assuming \eqref{d5internal} is small.

Using the trace of the energy momentum tensor for the D5-brane given in \eqref{d5trace} and the unwarped density in \eqref{pqdensity} gives \eqref{d5internal} to be
\begin{align}
2 \kappa_{10}^2 T_5\mathcal{H}^{-1/2}\left ( \frac{1}{2} \mathcal{H}^{-1}\tilde{\rho}_{3}^{pD5} -  \mathcal{H}^{-3/2}\tilde{\rho}_3^{pqD5} \right ). \label{d5internalsource}
\end{align}
Now that we have the warp factor, we can now check the size of each term individually, and check that they are small relative to the LHS of \eqref{bianchid5}, which scales as 
$
\tilde{\nabla}^2 \Phi_- = \mathcal{O}(\delta/{r_{\text{min}}^2} ) $ in the small $r$ region, and with $\Phi_- = e^{4A} - \alpha = \delta$ which is small. 
The first part of  \eqref{d5internalsource} is 
\begin{align}
\kappa_{10}^2 T_5\mathcal{H}^{-3/2} \tilde{\rho}_{3}^{pD5} &= \frac{25\pi}{32}\frac{1}{Ng_sl_s^4}\frac{r_{\text{min}}^7}{u^5\sin\theta_2}\delta(M_4^*).
\end{align}
For this term to be negligable, we need to impose
\begin{align}
\gamma &\ll 1 \label{rtiplim}
\\
\text{where }\gamma \equiv \frac{\beta}{\delta},  \qquad \beta &\equiv \frac{25\pi}{32}\frac{1}{Ng_sl_s^4}\frac{r_{\text{min}}^9}{u^5\sin\theta_2}.
\end{align}

This may be possible to arrange for given values of $r_{\text{min}} \ll u$ in the SUGRA limit of large $N$. Note that $l_s \ll u$ is required for the curvature of the WRC geometry to not be too large - necessary for the SUGRA approximation. We also work in the perturbative regime of small $g_s$. There is no restriction on the size of $r_{\text{min}}/l_s$, as $r_{\text{min}}$ isn't a curvature term, it's just a coordinate distance in the WRC, and is set by the minimum of the potential. We see that as long as we stay away from $\theta_2 = 0$, \eqref{rtiplim} can be satisfied for a suitable potential. Note that this condition \eqref{rtiplim} is more stringent than the condition from the backreaction on the warp factor.

The data from Table~\ref{tab:parameters}, together with the values of $p$ in \eqref{pd5} and $q$ in \eqref{qgut} gives
\begin{align}
\gamma &= \frac{25\pi}{ 32\delta}\frac{1}{Ng_sl_s^4}\frac{r_{\text{min}}^9}{u^5\sin\theta_2}
\\
&\approx 8\times 10^{-8} \ll 1
\end{align}
as required. 

The second term in \eqref{d5internalsource} is
\begin{align}
 -2 \kappa_{10}^2T_5 \mathcal{H}^{-2}  \tilde{\rho}_3^{pqD5} &= -\frac{25\pi}{32} \frac{1}{Ng_sl_s^4}\frac{r_{\text{min}}^7}{u^5\sin\theta_2} \left [48 \pi^{1/2}\frac{q}{\sqrt{Ng_s}}\right ] \delta(M_4^*)
\end{align}
which is small compared to $\delta/{r_{\text{min}}^2}$ if 
\begin{align}
q  &\ll  10^{-2}\gamma^{-1}{\sqrt{Ng_s}}, \label{qlim}
\end{align}
which for the $q$ in \eqref{qgut} requires
\begin{align}
1.2 \times 10^{-15} \frac{N^{3/2}}{g_s^{1/2} \delta B} \left (\frac{u}{l_s}\right )^4 \left (\frac{u}{r_{\text{min}}}\right )^2 &\ll  10^{-2}\gamma^{-1}{\sqrt{Ng_s}}.
\end{align}
Substituing $\gamma=\beta/\delta$ and tidying up means we require
\begin{align}
\frac{1}{B g_s^2 \delta^2 \sin^2\theta_2}\frac{u}{l_s}\left (\frac{r_{\text{min}}}{l_s}\right )^7 \ll 3.4\times 10^{12}
\end{align}
which is independent of $N$. Putting in the data from Table~\ref{tab:parameters} gives
\begin{align}
\frac{1}{B g_s^2 \delta^2 \sin^2\theta_2}\frac{u}{l_s}\left (\frac{r_{\text{min}}}{l_s}\right )^7 &\sim 5\times 10^{6} \ll 3.4\times 10^{12}.
\end{align}

It's interesting to note that this condition on $q$ from backreaction is less restrictive than that coming from our approximation for $\mathcal{F}$ in \eqref{qfapprox}, which we used to get a Planckian decay constant. This was shown to be satisfied for our data in \eqref{fapproxcorrect}.
It seems that setting a Planck scale decay constant together with a hierarchically smaller GUT scale of inflation is a more delicate procedure than maintaining control over the backreaction from the flux of a wrapped brane.

\section{Corrections from the 4D Ricci Scalar} \label{ricciappend}
We now consider the effect of a non-negligible 4D Ricci scalar $\mathcal{R}_4$. The solution to the full Poisson equation \eqref{poissonphi-} is denoted $\Phi_-$. For a 4D quasi-de Sitter spacetime, we have that
\begin{align}
{\tilde{\nabla}}^{2}  \Phi_- = \mathcal{R}_4 \approx  12H^2 \approx \frac{4V}{M_p^2} \approx \varphi(y) + \lambda \Phi_- . \label{phipot1}
\end{align}  
By the Friedmann equation, $H^2 = V/(3M_p^2)$, where we have seperated the $\Phi_-$ dependence from the rest of the potential. The potential $V$, and hence $\varphi(y)$ and $\lambda$, depend on the choice of probe D-brane.

For a probe D3-brane, we have $\varphi(y)= V_0$, a constant and $\lambda=4T_3/M_p^2$. This can be solved exactly,   
 in the region $r_{\text{min}} \ll r \ll r_{\text{UV}}$, in which case the geometry of the SC is relevant \cite{2010JHEP...06..072B}. The solutions are modified Bessel functions with argument $x \equiv \sqrt{\lambda} r$. When these are expanded for small $x$, a mass term appears as the leading curvature correction, leading to the eta problem as seen from the 10D supergravity perspective. 

Note that in \cite{Gregory:2011cd} the Ricci scalar term was omitted from the equation of motion for $\Phi_-$. However, it was argued that a mass term should be added to the final potential. The origin of this mass term is this curvature.

We now consider the case of a probe D5-brane moving in the WRC throat, in which case  $\varphi(y)$ is not constant. The solution to \eqref{phipot1} in this case can be derived using the expansion method developed in \cite{2010JHEP...06..072B}, which was originally used for the case where the perturbations to $\Phi_-$ come from fluxes and curvature at leading order. The expansion takes the form
\begin{align}
\Phi_- &= \sum_{n=0}^{\infty}\Phi_-^{[n]} \label{expandphi}
\\
\text{where } \Phi_-^{[0]}(y) &=  \overline{\Phi}_-(y)+ \Phi_{h}(y) \label{greenint}
\\
\text{and } \Phi_-^{[n]}(y) &= \lambda\int \sqrt{\tilde{g}_6'}d^6y' G(y;y')\Phi_-^{[n-1]}(y'),\end{align}
where $\Phi_{h}(y)$ is the solution to the homogeneous Laplace equation, and $\overline{\Phi}_-(y)$ is sourced by $\varphi(y)$ via
\begin{align}
\overline{\Phi}_-(y)  \equiv \int \sqrt{\tilde{g}_6'}d^6y' G(y;y')\varphi(y'). \label{inhom0}
\end{align}
In the above, $G$ is the Green's function satisfying
\begin{align}
{\tilde{\nabla}}^{2}_y G(y;y') = \frac{\delta(y-y')}{\sqrt{\tilde{g}_6'}}. \label{Greenseqn}
\end{align}

The expansion \eqref{expandphi} can be truncated if $\lambda r^2 < 1$, in which case the leading order term is from $\Phi_-^{[0]}$. 
We will now calculate the  $\overline{\Phi}_-(y)$ term for a D5-brane in the WRC. We will find that the leading correction from the Ricci scalar is subdominant to $\varphi(y)$ in the small $\rho$ limit, contributing a term of only at order $\rho^4$, which is small. 

Using our chosen $p$ and $q$ we get 
\begin{align}
\varphi(y) &= \frac{4}{M_p^2}4\pi u^2pT_5\mathcal{H}^{-1/2}  \approx \frac{180}{u^2}\rho^2
\\
\lambda &=  \frac{4}{M_p^2}4 \pi^2 l_s^2 T_5 p q \approx 4 \frac{M_{\text{GUT}}^4}{\delta M_p^2} \approx \frac{0.2}{\delta u^2} \approx \frac{20}{ u^2}
\end{align}
where we've used that $M_{\text{GUT}} = 4\times 10^{-3}M_p$, and 
\begin{align}
M_p^2 &= \frac{Nu^2}{2(2\pi)^4g_s l_s^4} \approx\frac{ 2 \times 10^{8}}{u^2}
\end{align}
for our data.
Then we have
\begin{align}
\lambda M_p^2/4 \approx 100 M_{\text{GUT}}^4. \label{lambdamp}
\end{align}
We note that in the potential there is the term
\begin{align}
\frac{M_p^2}{4}\varphi(\rho) \approx M_{\text{GUT}}^4 880 \rho^2 \label{vpot}
\end{align}
coming from the non-cancellation of DBI and CS terms.

The expansion of ${\Phi_-}(y)$ in  \eqref{expandphi} can be truncated if $\lambda r^2 \ll 1$. 
In our case, we are interested in probing the $r \sim r_{\text{min}} \sim u/50$
 region, in which case $\lambda r^2 \approx 20/2500 \approx 8\times 10^{-3} \ll 1$. Then the leading order term is $ \Phi_-^{[0]}$. We have chosen $\Phi_h$, and so now we need to calculate $\overline{\Phi}_-$. To compute this we need to calculate the Green's function and integrate \eqref{inhom0} to find the leading order small $\rho$ behaviour of $\overline{\Phi}_-$.

We can calculate the Green's function using the eigenfunctions, $Y_{L}(Z)$, of the Laplacian on $T^{1,1}$. 
The delta function on the RC splits into a radial delta function and the delta function on $T^{1,1}$,
\begin{align}
\delta(y-y') &= \delta(r-r')\prod_{i=1}^{i=5}\delta(Z_i-Z_i').
\end{align}
The delta function on the angular parts can be expanded in the $Y_{L}(Z)$ 
\begin{align}
\frac{\prod_{i=1}^{i=5}\delta(Z_i-Z_i')}{\sqrt{\tilde{g}_5}} = \sum_L Y_{L}(Z_i)Y^*_{L}(Z_i')
\end{align}
which have the conventional normalisation
\begin{align}
\int d^5Z_i \sqrt{\tilde{g}_5} Y_L^*(Z_i)Y_{L'}(Z_i) = \delta_{LL'} \label{normy} 
\end{align}
where  
\begin{align}
\sqrt{\tilde{g}_5} \equiv \frac{\sqrt{\tilde{g}_6}}{\sqrt{\tilde{g}_r}} = \frac{r^3(r^2+6u^2)}{\sqrt{\tilde{g}_r}}\frac{\sin\theta_1\sin\theta_2}{108} =  \frac{\sin\theta_1\sin\theta_2}{108}.
\end{align}

The $Y_L$ are given by 
\begin{align}
Y_L(Z_i) &= J_{l_1,m_1,R}(\theta_1)J_{l_2,m_2,R}(\theta_2)e^{i(m_1\phi_1 + m_2\phi_2  + R\psi/2)}
\\
\begin{split}
\text{where }  J^{\Upsilon}_{l_i,m_i,R}(\theta_i) &= N_{\Upsilon} (\sin \theta_i)^{m_i}\left (\cot \left (\frac{\theta_i}{2}\right )\right)^{R/2} \times
\\
& {}_2F_1\left (-l_i + m_i, 1 + l_i + m_i; 1 + m_i - R/2; \sin ^2 \left (\frac{\theta_i}{2}\right )\right )
\end{split}
\\
\begin{split}
\text{and }  J^{\Omega}_{l_i,m_i,R}(\theta_i) &= N_{\Omega} (\sin \theta_i)^{R/2}\left (\cot \left (\frac{\theta_i}{2}\right )\right)^{m_i} \times
\\
& {}_2F_1\left (-l_i + R/2, 1 + l_i + R/2; 1 + R/2 - m_i; \sin ^2 \left (\frac{\theta_i}{2}\right )\right ).
\end{split}
\end{align}
The $\Upsilon$ solution is regular for $m_i \geq R/2$, while $\Omega$ is regular for $m_i \leq R/2$. $N_{\Upsilon}$ and $N_{\Omega}$ impose the normalisation \eqref{normy}.

In order to obtain single-valued regular functions, the charges must satisfy
\begin{itemize}
\I $l_1$ \& $l_2$ both integers or both half-integers
\I $m_1 \in \{ -l_1,...,l_1 \}$ and $m_2 \in \{ -l_2,...,l_2 \}$
\I $R \in \mathbb{Z}$ and $\dfrac{R}{2} \in \{ -l_1,...,l_1 \}$ and $\dfrac{R}{2} \in \{ -l_2,...,l_2 \}$.
\end{itemize}

The Green's function can be expanded in these $Y_L$, as 
\begin{align}
G(y;y') = \sum_L G_L(r;r')Y_{L}(Z_i)Y^*_{L}(Z_i'). \label{gleqn}
\end{align}
Then the $\overline{\Phi}_-(y)$ term is 
\begin{align}
\overline{\Phi}_-(y) &= \sum_{L}Y_{L}(Z_i)  \int d^5Z_i' \frac{\sin\theta_1'\sin\theta_2'}{108}Y^*_{L}(Z_i')\int\sqrt{\tilde{g_r}'}dr'  G_{L}(r;r')\varphi(r'). 
\label{integralform}
\end{align}
We now prove that if $\varphi(y)$ has no angular dependence, then $\overline{\Phi}_-(y)$ only has a contribution from the singlet $L=(0,0,0,0,0)$, due to the vanishing of the angular integral
\begin{align}
\int d^5Z_i' \frac{\sin\theta_1'\sin\theta_2'}{108}Y^*_{L}(Z_i') \label{angularint}
\end{align}
 for other $L$.
 
To see this, note that because $0 \leq \psi < 4\pi$, we must have $R=0$, for $\int_0 ^{4\pi} e^{iR\psi/2} \neq 0$. For $R=0$, we must have that $l_1, l_2$ are both integers. Hence $m_1,m_2$ must both be integers.

Similarly, in order for $\int_0 ^{2\pi} e^{im_i\phi_i} \neq 0$, for $m_i$ integers, we must have $m_1 = 0 =m_2$. For $m_i=R=0$, the $\Upsilon$ and $\Omega$ solutions coincide, so we drop these labels. The form of the $\theta_i$ dependence will simplify to the Legendre polynomials $P_{l_i}$
\begin{align}
J_{l_i,0,0} &= N{}_2F_1\left (-l_i, 1 + l_i; 1; \sin ^2 \left (\frac{\theta_i}{2}\right )\right )
\\ &= N P_{l_i}(\cos \theta_i).
\end{align}

To see this, note that the Jacobi Polynomials $P_{n}^{(\alpha,\beta)}(x)$ are defined in terms of the hypergeometric function by 
\begin{align}
P_{n}^{(\alpha,\beta)}(z) &= \frac{(\alpha +1)_{n}}{n!} {}_2F_1\left (-n, 1 + n + \alpha + \beta
; 1 + \alpha; \frac{1}{2}(1 - z)\right )
\end{align}
where $(\alpha +1)_{n}$ is the rising Pochhammer symbol. We have
the special case $n=l_i, \alpha = 0 = \beta$, and $z = \cos \theta_i$, for which the Jacobi polynmials reduce to the Legendre polynomials
\begin{align}
P_{n}^{(0,0)}(z) = P_n(z)
\end{align}
and the Pochhammer symbol is $(1)_n = n!$, so that
\begin{align}
J_{l_i,0,0} &= NP_{l_i}(\cos \theta_i).
\end{align}
We then evaluate the integral using the orthogonality of Legendre polynomials, noting that $P_0(z) = 1$, and making the substitution $z=\cos \theta_i$
\begin{align}
\int_0^{\pi} d \theta_i  \sin\theta_i P_{l_i}(\cos \theta_i) &= \int_{-1}^{1} P_{l_i}(z) d z 
\\
&= \delta_{0l_i}
\end{align}
which vanishes unless $l_i=0$. Thus the only contribution to \eqref{angularint} is from the singlet $L = (0,0,0,0,0)$.
This has a constant eigenfunction $Y_{\{0\}}(Z_i) = \alpha$, where $\alpha$ is set by the normalization \eqref{normy}, giving
\begin{align}
\int d^5Z_i \frac{\sin\theta_1\sin\theta_2}{108}|\alpha|^2 = 1
\\
\Rightarrow |\alpha| = \sqrt{\frac{27}{16 \pi^3}}.
\end{align}
Now looking at \eqref{integralform}, the only non-vanishing contribution comes from the $L=\{0\}$ term, which has angular part
\begin{align}
Y_{\{0\}}(Z_i)  \int d^5Z_i' \frac{\sin\theta_1'\sin\theta_2'}{108}Y^*_{\{0\}}(Z_i') = |\alpha| ^2 * |\alpha| ^{-2} = 1.
\end{align}
We then just need to work out the radial integral
\begin{align}
\overline{\Phi}_-(y) &= \int\sqrt{\tilde{g_r}'}dr'  G_{\{0\}}(r;r')\varphi(r').
\\
&=  (3u)^{6}\int\sqrt{\tilde{g_{\rho}}'}d{\rho}'  G_{\{0\}}({\rho};{\rho}')\varphi({\rho}').
\end{align}

For each $G_L(r;r')$ in \eqref{gleqn}, we have the radial equation on the RC
\begin{align}
\begin{split}
&\frac{1}{r^3(r^2 + 6u^2)}\p_r(r^3(r^2 + 9u^2)\p_rG_L) 
\\
&- \left [ \frac{6(l_1(l_1+1) - R^2/4)}{r^2} + \frac{6(l_2(l_2+1) - R^2/4)}{r^2 + 6u^2}  + \frac{9R^2/4}{\kappa r^2}\right ]G_L = \frac{\delta(r - r')}{r^3(r^2 + 6u^2)}
\end{split}
\end{align}
which for dimensionless $\rho$ becomes
\begin{align}
\begin{split}
&\frac{1}{\rho^3(\rho^2 + 2/3)}\p_{\rho}(\rho^3(\rho^2 + 1)\p_{\rho}G_L) 
\\
&- \left [ \frac{6(l_1(l_1+1) - R^2/4)}{\rho^2} + \frac{6(l_2(l_2+1) - R^2/4)}{\rho^2 + 2/3}  + \frac{9R^2/4}{\kappa \rho^2}\right ]G_L = \frac{\delta(\rho - x)}{(3u)^4\rho^3(\rho^2 + 2/3)}. \label{rhogreen}
\end{split}
\end{align}
Here we've written $\rho'$ as $x$, for clarity of variables in the following. The good news about \eqref{rhogreen} is that we can solve it exactly for $L=\{0\}$ on the whole of the RC, including the region of small $\rho$. For $L=\{0\}$ we just have
\begin{align}
 \p_{\rho}^2 G_{\{0\}} + \left ( \frac{5\rho + 3\rho^{-1}}{\rho^2 + 1}\right )\p_{\rho}G_{\{0\}} &= \frac{\delta(\rho - x)}{(3u)^4}.
 \end{align} 
The solution is
\begin{align}
G_{\{0\}}(\rho;x) &= \frac{1}{(3u)^4} 
\begin{cases}
-(2x^2)^{-1} - \log x + \frac{1}{2}\log (x^2 +1)  & \text{if } \rho \leq x
\\
-(2\rho^2)^{-1} - \log \rho^{} + \frac{1}{2}\log (\rho^2 +1)  & \text{if } \rho \geq x.
\end{cases}
\end{align}
Viewed as a function of $\rho$, the $\rho \leq x$ part of the solution is just a constant, fixed by continuity of $G_{\{0\}}$, while the $\rho \geq x$ part is the non-constant solution to the homogeneous equation, regular at infinity.

Now we do the Green's function integral for the $L=\{0\}$ mode. The integral is done in two pieces, the first is for $x \leq \rho$
\begin{align}
\begin{split}
\int_0^{\rho } x^5 \left(x^2+\frac{2}{3}\right) \left(-\frac{1}{2 \rho ^2}+\frac{1}{2} \log \left(\rho ^2+1\right)-\log \rho \right) \, dx
\\
= -\frac{1}{144} \rho ^4 \left(9 \rho ^2+8\right) \left[2 \rho ^2 \log \rho -\rho ^2 \log \left(\rho ^2+1\right)+1\right]
\end{split}
\end{align}
the second is for 
$x \geq \rho$
\begin{align}
\begin{split}
&\int_{\rho }^{1/3} x^5 \left(x^2+\frac{2}{3}\right) \left(-\frac{1}{2 x^2}+\frac{1}{2} \log \left(x^2+1\right)-\log x\right) \, dx
\\
= &\frac{1}{23328}\bigg[ 162  \log \left(9 \left(\rho ^2+1\right)\right)+162 \left(9 \rho ^2+8\right) \rho ^6 \left(2 \log \rho -\log \left(\rho ^2+1\right)\right) 
\\ 
& +81 \left(18 \rho ^4+25 \rho ^2-2\right) \rho ^2-9-160 \log (10) \bigg ].
\end{split}
\end{align}
Putting these together gives the exact result
\begin{align}
\begin{split}
\overline{\Phi}_-(y)  = (3u)^6\int\sqrt{\tilde{g_{\rho}}'}dx  G_{\{0\}}(\rho;x)\frac{180}{u^2}x^2
=
\frac{5}{72}\bigg [&81 \left(9 \rho ^2-2\right) \rho ^2+162 \log \left(9 \left(\rho ^2+1\right)\right)
\\
&-9-160 \log (10)\bigg]
\end{split}
\end{align}
which, for small $\rho$ has leading order behaviour
\begin{align}
\overline{\Phi}_-(y) 
&\approx
-1.4905  + 45 \rho ^4 + \frac{15 }{4}\rho ^6 + \mathcal{O}(\rho^{8}).
\end{align}
Using \eqref{lambdamp}, we see that in the potential will appear the term $100M_{GUT}^445 \rho ^4 = M_{GUT}^4 4500\rho ^4$, as in \eqref{potcorrected}. This doesn't contribute quantitatively in the small $\rho$ limit.

The result that only the $L=\{0\}$ mode contributes to $\overline{\Phi}_-(y)$ means that $\overline{\Phi}_-(y)= \overline{\Phi}_-(r)$ is purely radial. 
If the homogeneous solution $\Phi_h(y)$ is also purely radial, then $\Phi_-^{[0]}(y)$ is also purely radial. The results of the above can then be applied so that we would have all $\Phi_-^{[n]}(y)$ purely radial since the integrals over the $Y_L$ would vanish for $L\neq \{0\}$. 

However, in our case, we have angular dependence in $\Phi_h(y)$ for the form of the Natural Inflation potential, which would induce angular dependence in $\Phi_-^{[n]}(y)$ for higher $n$.

\section{Wrapped D5-branes and the $b$-Axion} \label{appendix:d5b2}
Wrapped D5-branes can source a potential for the NS-NS axion, $b$, which arises from the integral of the 2-form field $B_2$ over a 2-cycle in a type IIB compactification. The application of this to inflation was investigated in the original axion monodromy models, for example in \cite{Silverstein:2008sg}, where the couplings of the axion are generated just from the DBI part of the D5-brane action. 
Our model of natural inflation involves a wrapped D5-brane giving rise to a potential for this field, and so we should consider whether these $b$-axions are relevant to inflation. 

In our case, we can choose to turn on a worldvolume $B_2$ flux of strength $b$ along the wrapped 2-cycle so that its pullback has the following non-zero components $P_6[B_2]_{\theta_1 \phi_1}   = \frac{b}{2} \sin \theta_1 $. Then a potential for $b$ enters via the function ${\mathcal{F}} $ in \eqref{wd5}, with  $2\pi\alpha' q$ shifted to $2\pi\alpha' q+b$. The same analysis leading to natural inflation with a Planckian decay constant follows as long as we make the choice $b \lae 2\pi\alpha' q$, together with the upper bound on $q$ coming from \eqref{qfapprox}. In contrast, models of axion monodromy inflation assume that  $ b \gg l $ initially, where $l$ is the size of the wrapped $S^2$. In this way the $b$-axion acquires  a linear potential effectively from expanding the ${\cal F}$ term leading to large field inflation. In our model this would mean taking the $2\pi\alpha’ q+b$ term to be the dominant factor inside ${\cal F} $ which is the opposite of what we have assumed. Ultimately this is a choice of initial condition on the  value of $b$. Our choice of  initial condition is that $b  \leq  2\pi\alpha’ q$  so that $2\pi\alpha’ q+b$ is  still sub-dominant to the warp factor term in ${\cal F} $ when the brane is near the tip. As such $b$ will not play a role in generating significant inflation and we may ignore it.

Finally, in more recent axion monodromy models \cite{ Marchesano2014} the monodromy is induced not by the DBI part of the wrapped D5-brane action, instead through background 3-form fluxes coupling to the $b$-axion associated with $B_2$ in the CS part of the action. However, our model is constructed using the WRC in which SUSY preserving 3-form fluxes are absent and so no such monodromy is induced either. Turning on such fluxes may be of interest in discussing moduli stabilisation mechanisms within the WRC which is something that we would like to investigate further. 

The conclusion of this appendix is that although the inclusion of wrapped D5-branes in our model could source potentials for $b$-axions, and hence complicate the inflationary picture, our assumed initial conditions on the size of $b$ ensure that it will not contribute towards the inflationary dynamics. However, more complicated models could relax these initial conditions.

\bibliographystyle{JHEP}

\bibliography{wrc_NI2}

\end{document}